\documentclass[english,aps,prl,floatfix,superscriptaddress]{revtex4}
\usepackage[T1]{fontenc}
\usepackage{amssymb}
\usepackage{amsmath}
\usepackage{graphicx}
\usepackage{float}
\usepackage{esint}
\usepackage{babel}
\usepackage[normalem]{ulem}

\usepackage{subfig}
\newcommand{\lyxdot}{.}

\begin{document}

\title{Chimera patterns in conservative systems and ultracold atoms with mediated nonlocal hopping}

\author{Hon Wai Lau}
\affiliation{Institute for Quantum Science and Technology and Department of Physics and Astronomy, University of Calgary, Calgary, Alberta, Canada T2N 1N4}
\affiliation{Complexity Science Group, Department of Physics and Astronomy, University of Calgary, Canada T2N 1N4}

\author{J\"{o}rn Davidsen}
\affiliation{Complexity Science Group, Department of Physics and Astronomy, University of Calgary, Canada T2N 1N4}
\affiliation{Hotchkiss Brain Institute, University of Calgary, Canada T2N 4N1}

\author{Christoph Simon}
\affiliation{Institute for Quantum Science and Technology and Department of Physics and Astronomy, University of Calgary, Calgary, Alberta, Canada T2N 1N4}
\affiliation{Hotchkiss Brain Institute, University of Calgary, Canada T2N 4N1}

\date{\today}

\keywords{Pattern formation, Nonlinear Dynamics, Atomic Physics, Condensed Matter Physics}

\begin{abstract}
Experimental realizations of chimera patterns, characterized by coexisting regions of phase coherence and incoherence, have so far only been achieved for non-conservative systems with dissipation. Moreover, theoretical studies of chimera patterns have also been limited either to the non-conservative case or to simplified models that describe the dynamics only in terms of a scalar phase field.
Here, we show for the first time explicitly that the formation of chimera patterns can also be observed in conservative Hamiltonian systems with nonlocal hopping in which both energy and particle number are conserved, and where the local phase and amplitude are non-separable even in the weak coupling regime.
Effective nonlocality can be realized in a physical system with only local coupling if different time scales exist, which we illustrate by a minimal conservative model with an additional mediating channel.
Finally, we show that chimera patterns should be observable in ultracold atomic systems:
Nonlocal spatial hopping over up to tens of lattice sites with independently tunable hopping strength and on-site nonlinearity can be implemented in a two-component Bose-Einstein condensate with a spin-dependent optical lattice, where the untrapped component serves as the matter-wave mediating field.
\end{abstract}

\flushbottom
\maketitle
\thispagestyle{empty}

\section*{Introduction}

Many interesting dynamical behaviors and physical phenomena observed in nature, including chaos, solitons, and many patterns in spatially extended systems, can only be modeled by nonlinear equations \cite{aranson_world_2002,kuramoto_chemical_1984,kapral_chemical_2012,cross_pattern_2009,panaggio_chimera_2015,pikovsky_synchronization:_2003,strogatz_nonlinear_2001,trombettoni_discrete_2001,eilbeck_discrete_1985}.
In this context, two fundamental nonlinear differential equations are the complex Ginzburg-Landau equation (CGLE) \cite{aranson_world_2002,kuramoto_chemical_1984} and the Gross-Pitaevskii equation (GPE) \cite{gross_structure_1961,gross_hydrodynamics_1963,pitaevskii_bose-einstein_2003}.
The CGLE corresponds to the normal form of any spatially extended system close to a Hopf bifurcation --- a critical point where a stationary system begins to oscillate \cite{kuramoto_chemical_1984,kapral_chemical_2012}.
It describes many physical systems phenomenologically, such as superconductivity and nonlinear waves \cite{landau_theory_1950,aranson_world_2002}.
Recently, the study of the CGLE with nonlocal diffusive coupling led to the discovery of an interesting dynamic pattern known as chimera states \cite{panaggio_chimera_2015,kuramoto_coexistence_2002,kuramoto_rotating_2003,shima_rotating_2004,kim_pattern_2004,martens_solvable_2010,motter_nonlinear_2010,gu_spiral_2013,lau_linked_2016,bera_chimera_2017,davidsen_symmetry-breaking_2018}
These states have been experimentally demonstrated in mechanical, chemical, electronic, and opto-electronic systems \cite{hagerstrom_experimental_2012,tinsley_chimera_2012,nkomo_chimera_2013, martens_chimera_2013,wickramasinghe_spatially_2013,schmidt_coexistence_2014, rosin_transient_2014,larger_virtual_2013,larger_laser_2015,kapitaniak_imperfect_2014} and also proposed to exist in metamaterials~\cite{lazarides_chimeras_2015}.
Chimera patterns are characterized by the coexisting of spatially localized regions of phase coherence and phase incoherence, even in a system with translational symmetry and starting from simple initial conditions.
So far, these patterns have been exclusively observed in experiments involving dissipative or non-conservative systems.

\begin{figure}
\begin{centering}
\includegraphics[width=0.6\columnwidth]{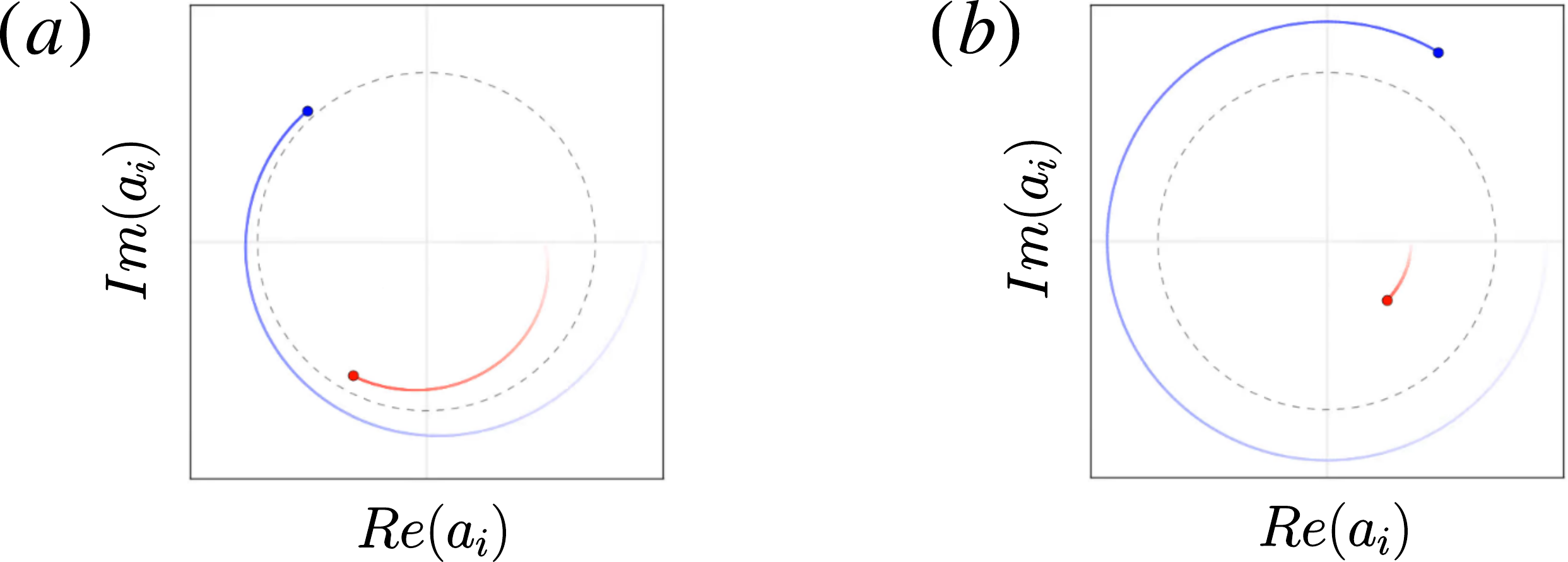}
\par\end{centering}

\centering{}\caption{\label{fig: illu-nonlinear-oscillators}
Illustration of the dynamics of two different types of oscillators in a two-dimensional phase space.
(a) A self-sustained oscillator with a limit cycle attractor. Trajectories near a limit cycle (represented by the dotted unit circle) move toward it. Energy disspation and driving are present such that two different initial states tend to the same asymptotic dynamics with the same oscillation frequency as time goes to infinity.
(b) A conservative nonlinear oscillator. Typically, the oscillation frequency depends on the initial condition. Since energy is conserved, trajectories corresponding to different initial energies remain separated at all times. 
}
\end{figure}

The GPE was originally derived as a mean-field description of interacting Bose-Einstein condensates (BECs) \cite{dalfovo_theory_1999,leggett_bose-einstein_2001,anglin_boseeinstein_2002,pethick_bose-einstein_2008}.
It is the Schr{\"o}dinger equation but includes an extra nonlinearity.
In contrast to the typical regime of the CGLE, GPE describes a conservative system that locally behaves like an undamped non-linear oscillator (see Fig. \ref{fig: illu-nonlinear-oscillators}).
Both field equations have global phase symmetry, a third-order nonlinearity, and become equivalent in certain limits \cite{aranson_world_2002,cross_pattern_2009}.
Given these similarities, it is natural to ask whether chimera patterns could also be observed for systems described by the GPE, i.e., in the conservative case.
Here, we show that the answer is affirmative. 
Yet, unlike self-sustained oscillators with limit cycle attractors in dissipative systems, undamped oscillators in conservative systems can oscillate at any amplitude (see Fig. \ref{fig: illu-nonlinear-oscillators}).
As we show, this can lead to the emergence of chimera patterns with non-monotonically and strongly varying amplitudes, which has not been observed for limit-cycle oscillators.
As common for conservative systems, the initial conditions play a very important role for the formation of such chimera patterns, which can even arise from rather simple initial conditions.

We find that these chimera patterns exist in the two-component GPE, if there is nonlocal hopping (beyond nearest neighbor) with a new characteristic length scale $R$. 
Nonlocal descriptions are often conveniently employed for systems such as gravitational, electric, magnetic, and dipole interactions, despite locality being one of the basic principles of physics.
These descriptions are accurate when the mediating field is much faster than the dynamics of the particles, so that the mediating picture can be reduced to an effective particle-particle description with a nonlocal term.
Similar effective descriptions can be engineered by adding a mediating channel such as cavity-mediated global coupling \cite{ritsch_cold_2013}.
Moreover, the range of coupling may be tunable in certain systems such as those with nonlocal diffusive coupling \cite{tanaka_complex_2003} or long-range coupling mediated by light,  \cite{douglas_quantum_2015,gonzalez-tudela_subwavelength_2015} studied recently.
Tunable hopping \cite{de_vega_matter-wave_2008,navarrete-benlloch_simulating_2011,gonzalez-tudela_purely_2017,gonzalez-tudela_quantum_2017} has also recently been studied as an extension to Bose-Hubbard model (BHM), while the direct nearest-neighbor hopping is usually studied in the BHM \cite{jaksch_cold_1998,dutta_non-standard_2015,landig_quantum_2016}.
The physics of conservative systems that display both nonlocal hopping and nonlinearities are still rarely studied.

In this paper, we introduce three conservative Hamiltonian systems representing different levels of generality that give rise to chimera patterns.
The first mathematical and most general model is the nonlocal hopping model (NLHM), which can be considered to be the generalization of the discrete GPE with nonlocal hopping, or the mean-field of the BHM with nonlocal hopping.
Then we explain the origin of the effective nonlocal hopping from a minimal conservative model with only local coupling.
By attaching a fast mediating channel particles can be converted to, the adiabatic elimination of the fast channel results in the continuum NLHM.
Since we are aiming to identify a conservative physical system that can be well-described by the NLHM,
we finally propose a specific physical model based on a two-component BEC in a spin-dependent trap \cite{mckay_slow_2013} with Rabi oscillations \cite{scully_quantum_1997,gerry_introductory_2005}.
In this setup, the hopping originates from the spreading of wavefunctions in the mediating channel governed by a Schr\"odinger-like equation, so the spatial hopping of atoms is mediated by the matter-wave itself.
For this, we employ a mathematical formulation different from previous studies~\cite{de_vega_matter-wave_2008,navarrete-benlloch_simulating_2011}.
Since all three models preserve the same conservation properties and exhibit overall the same qualitative chimera behavior as we show, the NLHM can indeed be considered a good description for realistic systems in the non-perfect adiabatic elimination regime --- despite its generality.

Implementation of chimera patterns in BECs has the advantage of exploring both quantum and classical regimes, in addition to allowing flexible control of almost all parameters \cite{jaksch_cold_2005,morsch_dynamics_2006}.
For example, adjusting the particle density and the magnetic field near Feshbach resonances \cite{chin_feshbach_2010} can change both the rate of particle loss and the strength of nonlinear interactions.
The loss of ultracold atoms limits the lifetime (which can be critical in certain systems \cite{lau_proposal_2014}) and, in our proposal, limits the maximum observable range of hopping.
The difficulty in this implementation is to find the relevant conditions for non-trivial hopping ranges, which is detailed in the later parts of this paper.

\section*{Results}

\subsection*{Nonlocal hopping model \label{sec:Formulation-and-mechanism}}

\subsubsection*{Hamiltonian and dynamic equation}

The NLHM model is given by the Hamiltonian:
\begin{eqnarray}
\mathcal{H} = \mathcal{U}+\mathcal{P} = \frac{U}{2}\sum_{i}|a_{i}|^{4}-P\sum_{i,j}G_{ij}a_{i}^{*}a_{j}\label{eq: NLHM}
\end{eqnarray}
where $a_{i}=\sqrt{n_{i}}e^{i\theta_{i}}$ is a complex number representing the state of site $i$, such that $|a_i|$ is the amplitude, $n_{i}=|a_{i}|^{2}$ is the number of particles or density, and $\theta_{i}$ is the phase. 
$\mathcal{U}$ is the nonlinear energy with the on-site nonlinear interaction $U$, and $\mathcal{P}$ is the hopping energy with the hopping strength $P$.
$G_{ij}$ is the hopping kernel describing the hopping from site $\mathbf{r}_{j}$ to $\mathbf{r}_{i}$, with $G_{ij}=G_{ji}$.
Typically, $G_{ij}$ decreases as the distance $|\mathbf{r}_{j}-\mathbf{r}_{i}|$ increases and may be characterized by a hopping range $R$.
For sufficiently small $R$, the hopping effectively becomes nearest neighbor.
This Hamiltonian conserves both the energy and the particle number $N=\sum_{i}n_{i}$.
It can also be expressed using the canonical coordinate and momentum $\{q_{i},p_{i}\}$, as well as action and angle variable $\{n_{i},\theta_{i}\}$ (see SM sec. S1).
Note that the hopping term is quadratic $a_{i}^{*}a_{j}$ in the Hamiltonian, which is different from the usual quartic term of a particle-particle interaction $n_{i}n_{j}$ for, say, the Coulomb interaction.
Therefore, the corresponding dynamical equation contains
the lowest order on-site nonlinearity and the nonlocal linear hopping term:
\begin{equation}
i\hbar\dot{a}_{i}=U|a_{i}|^{2}a_{i}-P\sum_{j}G_{ij}a_{j}\label{eq: NLHM-dynamic}
\end{equation}
where $\hbar$ is the Planck constant, which we can set to $\hbar=1$ without loss of generality by rescaling time. Note that this equation is the mean-field equation of the BHM with nonlocal hopping \cite{de_vega_matter-wave_2008,navarrete-benlloch_simulating_2011}. 
Moreover, the nearest-neighbor variation of this equation is the discrete GPE \cite{trombettoni_discrete_2001} and the non-spatial variation
is the discrete self-trapping equation \cite{eilbeck_discrete_1985}.

The dynamic equation of the NLHM can be rewritten in a dimensionless form using the rescaling $a_{i} \to a_{i}/\sqrt{n_{0}}$, $t \to (Un_{0}/\hbar)t$, and $P \to P/(Un_{0})$ where $n_{0}$ is the average number of particles per site.
The equation becomes: $i\dot{a}_{i}(t)=|a_{i}|^{2}a_{i}-P\sum_{j}G_{ij}a_{j}$, which depends only on the control parameters of rescaled hopping strength $P$ and rescaled hopping radius $R$.
Alternatively, Eq. (\ref{eq: NLHM-dynamic}) can be written in terms of $\theta_{i}(t)$ and $n_{i}(t)$ as
\begin{subequations}\label{eq: phase-density-equation}
\begin{eqnarray}
\dot{\theta}_{i}(t) & = & Un_{i}-P\sum_{j}G_{ij}\sqrt{\frac{n_{i}}{n_{j}}}\cos(\theta_{j}-\theta_{i})\label{eq: phase-equation}\\
\dot{n}_{i}(t) & = & 2P\sum_{j}G_{ij}\sqrt{n_{i}n_{j}}\sin(\theta_{j}-\theta_{i})\label{eq: density-equation}
\end{eqnarray}
\end{subequations}
This explicitly shows that the evolution of the phase $\theta_{i}(t)$ depends on the density $n_{i}(t)$ of the oscillators and vice versa.
Even in the very weak hopping regime, they remain coupled to the lowest order.
For dissipative systems as illustrated in Fig. \ref{fig: illu-nonlinear-oscillators}a, if $\dot{n_i}\sim 0$ after dissipation in the weak coupling regime for all $i$, one can obtain a simplified phase dynamics.
This is generally not possible for the conservative case with constant energy since, in general, a large $n_i$ at some site $i$ has to be compensated by small $n_j$ at another site or sites to keep the energy constant.
This highlights the important role of these conditions for conservative systems in contrast to dissipative systems.
The dynamics of the NLHM can be found by solving Eq. (\ref{eq: NLHM-dynamic}) using standard numerical methods (see Methods), and the results for 1D and 2D are given in the following subsections.

\subsubsection*{Chimera patterns in 1D NLHM \label{sec:Chimera-state}}

\begin{figure}
\begin{centering}
\includegraphics[width=0.95\columnwidth]{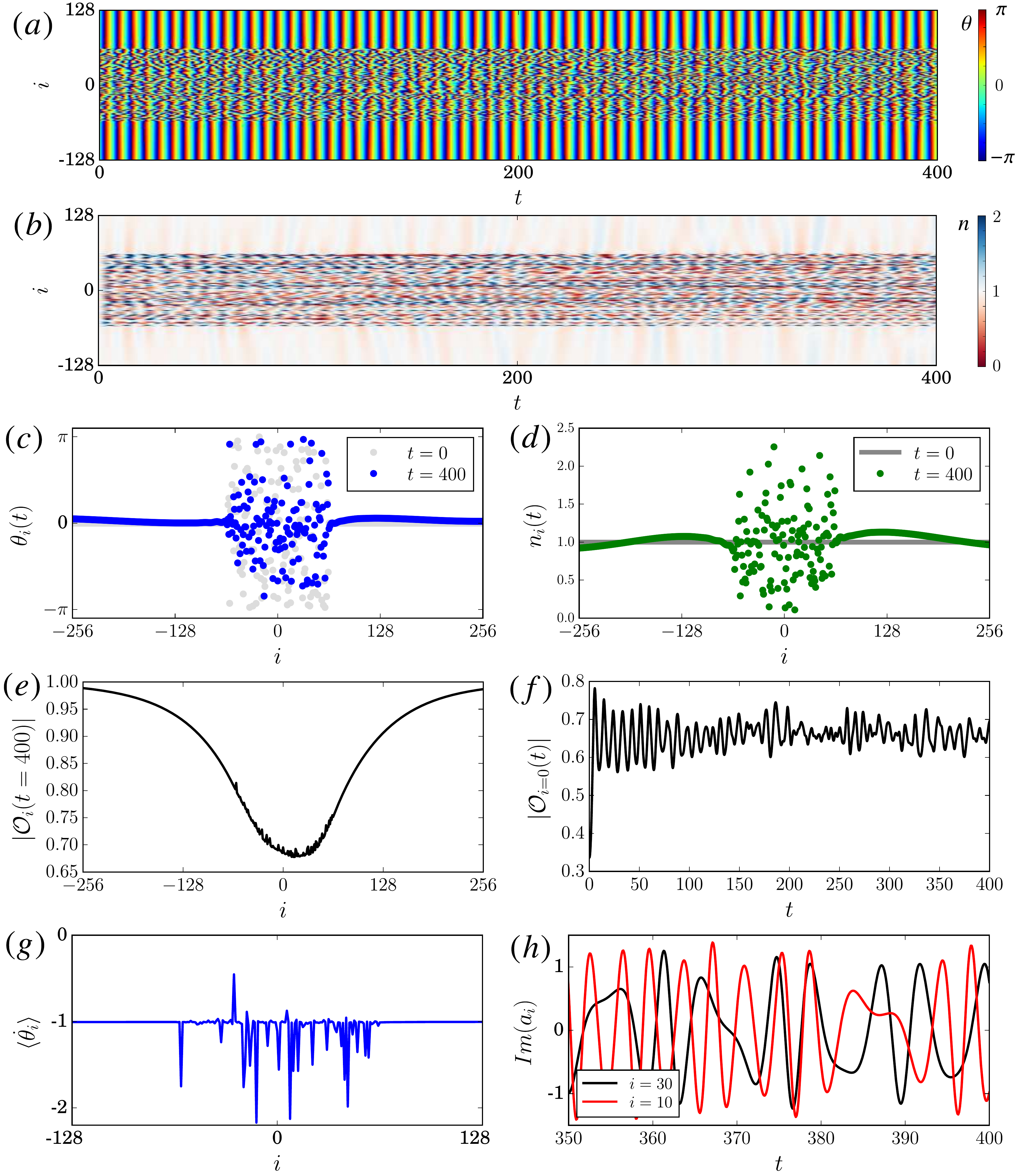}
\par\end{centering}

\centering{}\caption{\label{fig: nlhm1d-random}
NLHM in 1D with only random initial phases for oscillators.
(a) Space-time plot of the phase $\theta_i(t)$.
(b) Space-time plot of the $n_i(t)$.
(c) Snapshot of $\theta_i(t)$ at $t=0$ and $t=400$.
(d) Snapshot of $n_i(t)$ at $t=0$ and $t=400$.
(e) Plot of local order parameter $\mathcal{O}$ at $t=400$.
(f) Plot of local order parameter $\mathcal{O}$ at $x=0$ over time.
(g) Average angular frequency $\langle\dot{\theta}_i\rangle$ between $t=0$ to $t=400$.
(h) The oscillation $Im(a_i)$ of two oscillators near the center.
Parameters: $Un_{0}=1$, $P=0.2$, $R=64$, and number of lattice $L=2048$ with no-flux boundary condition and initial density $|a_i|^2=1$. Only the center region is shown for clarity. The hopping kernel $G_{ij}$ is given in Table \ref{tab: hopping-kernel}. Dimensionless units and $\hbar=1$ are used. 
}
\end{figure}

\begin{figure}
\begin{centering}
\includegraphics[width=0.95\columnwidth]{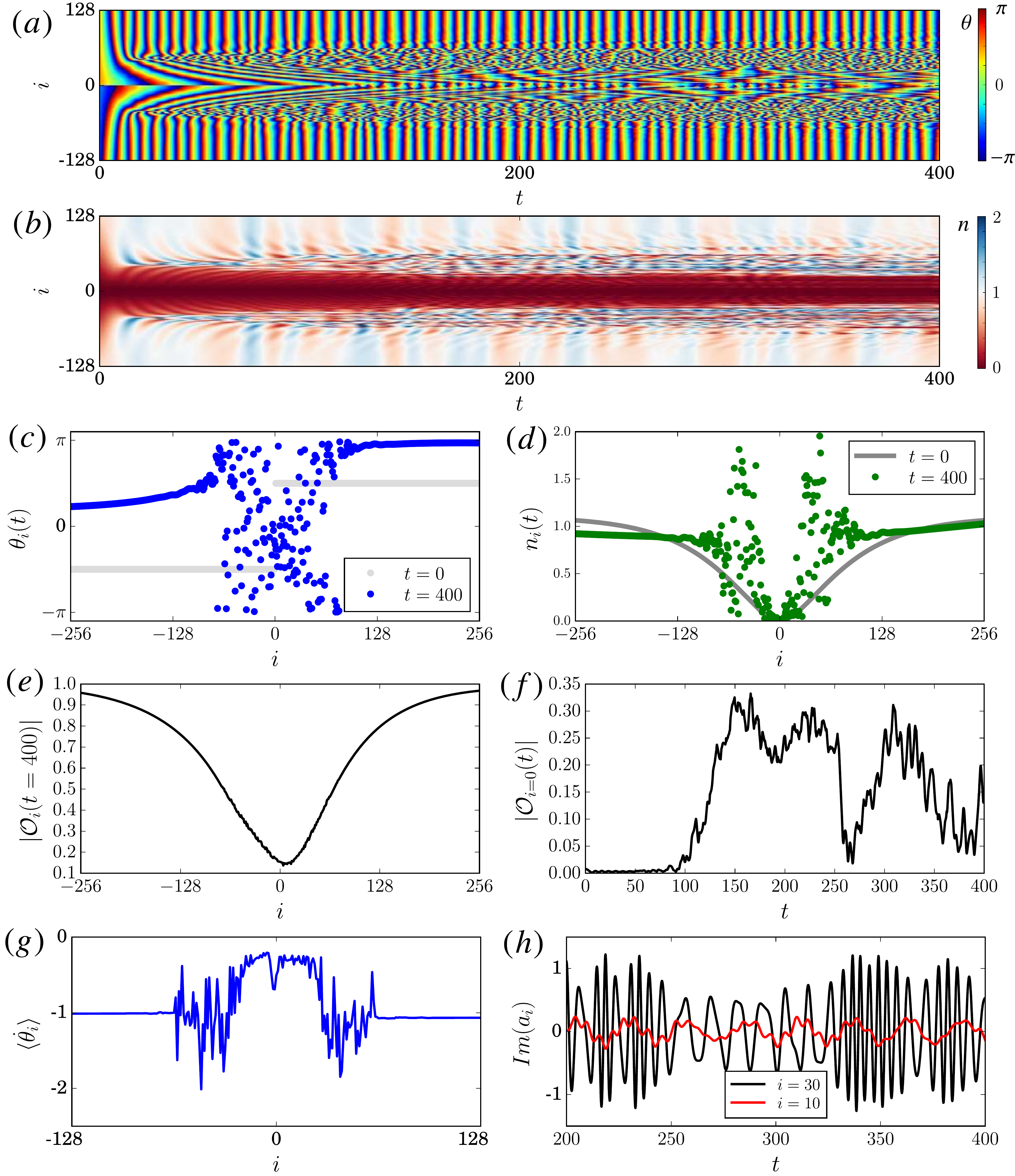}
\par\end{centering}

\centering{}\caption{\label{fig: nlhm1d-vortex-cross-section}
Similar to Fig. \ref{fig: nlhm1d-random}, but with initial zero density at the center and a phase flip as given in subfigure (c) and (d).
Same parameter as in Fig. \ref{fig: nlhm1d-random} with $N = \sum_i|a_i|^2 = L$.
}
\end{figure}

An often used initial condition for chimera patterns is a random phase field \cite{kuramoto_coexistence_2002,kim_pattern_2004,omelchenko_stationary_2012,maistrenko_chimera_2015}, in which chimera patterns can appear after a sufficiently long relaxation time. 
However, for NLHM, simulations show that the dynamics for such random initial conditions remains incoherent with no clear patterns over time. This is not unexpected since the \emph{spontaneous} emergence of \emph{persistent} patterns in spatially extended systems is typically tied to the notion of an attractor, which does not exist in our conservative model. 
Instead, incoherent and coherent regions --- and, thus, chimera states --- can sustain themselves over time as shown in Fig. \ref{fig: nlhm1d-random}(a)-(d) starting from initial conditions that are uniform with the exception of random phases (but not amplitudes or densities) in a small region. In particular, the time-averaged angular frequency $\langle\dot{\theta}_{i}\rangle$ as shown in Fig.~\ref{fig: nlhm1d-random}(g) is uniform in the coherent region and takes on a range of values in the incoherent region, thus, fulfilling the defining property of a chimera state.
In terms of the temporal evolution, even though $n_{i}$ is constant initially, the random phases immediately induce fluctuations in the density as shown in Fig. \ref{fig: nlhm1d-random}(b) (see animations of the simulations in SM) as expected based on Eq. (\ref{eq: phase-density-equation}). Such a behavior can not be captured by simplified phase models by construction.
To measure the coherence of the phase, we use the local order parameter $\mathcal{O}_i = \sum_j G_{ij} e^{i\theta_j}$ \cite{kuramoto_coexistence_2002}.
The magnitude $|\mathcal{O}_i| \sim 1$ when all phases $\theta_j$ are the same within the hopping range $R$. 
As shown in Fig. \ref{fig: nlhm1d-random}(e), $|\mathcal{O}_i|$ takes on a minimum near the center of the incoherent region as expected.
Moreover, the local order parameter does not converge but it keeps fluctuating as shown in Fig. \ref{fig: nlhm1d-random}(f) (see also Fig. \ref{fig: nlhm1d-vortex-cross-section}(f)) due to the conservative nature of the system, which prevents relaxation behavior typical for dissipative systems.
Figs. \ref{fig: nlhm1d-random}(a), \ref{fig: nlhm1d-random}(c), \ref{fig: nlhm1d-random}(e) and \ref{fig: nlhm1d-random}(g) also indicate that the incoherent region is not fully desynchronized for this initial condition.
A much stronger desynchronization can be obtained using an initial condition with both the phase and the amplitude random around the center region, see Fig. S2 in the SM.
Hence, the initial amplitude plays a significant role for the characteristics of the observed chimera patterns in conservative systems, while this is typically not the case for dissipative systems, where initial fluctuations in the amplitude tend to be damped away.
Another striking observation is that $\langle\dot{\theta}_{i}\rangle$ can behave non-monotonically across the incoherent core (see Fig.~\ref{fig: nlhm1d-random}(g) and also Fig.~\ref{fig: nlhm1d-vortex-cross-section}), whereas it typically changes monotonically with distance from the incoherent core in the dissipative case and in simplified phase models~\cite{kuramoto_coexistence_2002}. 
The incoherent dynamics of the oscillators can be observed in Fig. \ref{fig: nlhm1d-random}(h), where the trajectories of two oscillators inside the incoherent region are shown.
The specific value of the hopping strenght $P>0$ does not affect the chimera patterns qualitatively.
However, for uniform initial conditions in the amplitude, the fluctuations in the amplitude can decrease when $P$ decreases as shown in Fig. S3 in SM.

While this could suggest that a simple phase description is sufficient in some special cases, such a simplification is generally not possible as already discussed above.
Specifically, one distinctive feature of the NLHM is that the local phase oscillators can oscillate at any amplitude because of the lack of a limit cycle attractor.
This can be observed using an initial condition with different amplitudes. An example is given in Fig. \ref{fig: nlhm1d-vortex-cross-section} (c)-(d) where the initial density drops to zero and the phase changes by $\pi$ at the center (this is a cross-section of a vortex phase initial condition, see next section for more details).
As suggested by previous studies \cite{gu_spiral_2013,lau_linked_2016}, interesting chimera patterns can be formed spontaneously from such a regular initial condition.
Here, the local phase incoherence and local density fluctuation around the center increase over time as shown in Fig. \ref{fig: nlhm1d-vortex-cross-section}(a)-(b).
As Fig. \ref{fig: nlhm1d-vortex-cross-section}(h) shows, the instantaneous frequency of the oscillators near the center can also change significantly over time.
In particular, these oscillations have near zero amplitude as shown in Fig. \ref{fig: nlhm1d-vortex-cross-section}(b) and (h).
In contrast, for the corresponding chimera patterns formed in dissipative systems with self-sustained oscillators the oscillations typically evolve close to the limit cycles in the weak coupling regime \cite{kuramoto_rotating_2003}.

\subsubsection*{Chimera patterns in 2D NLHM \label{sec:Chimera-state2D}}

\begin{figure}
\begin{centering}
\includegraphics[width=0.9\columnwidth]{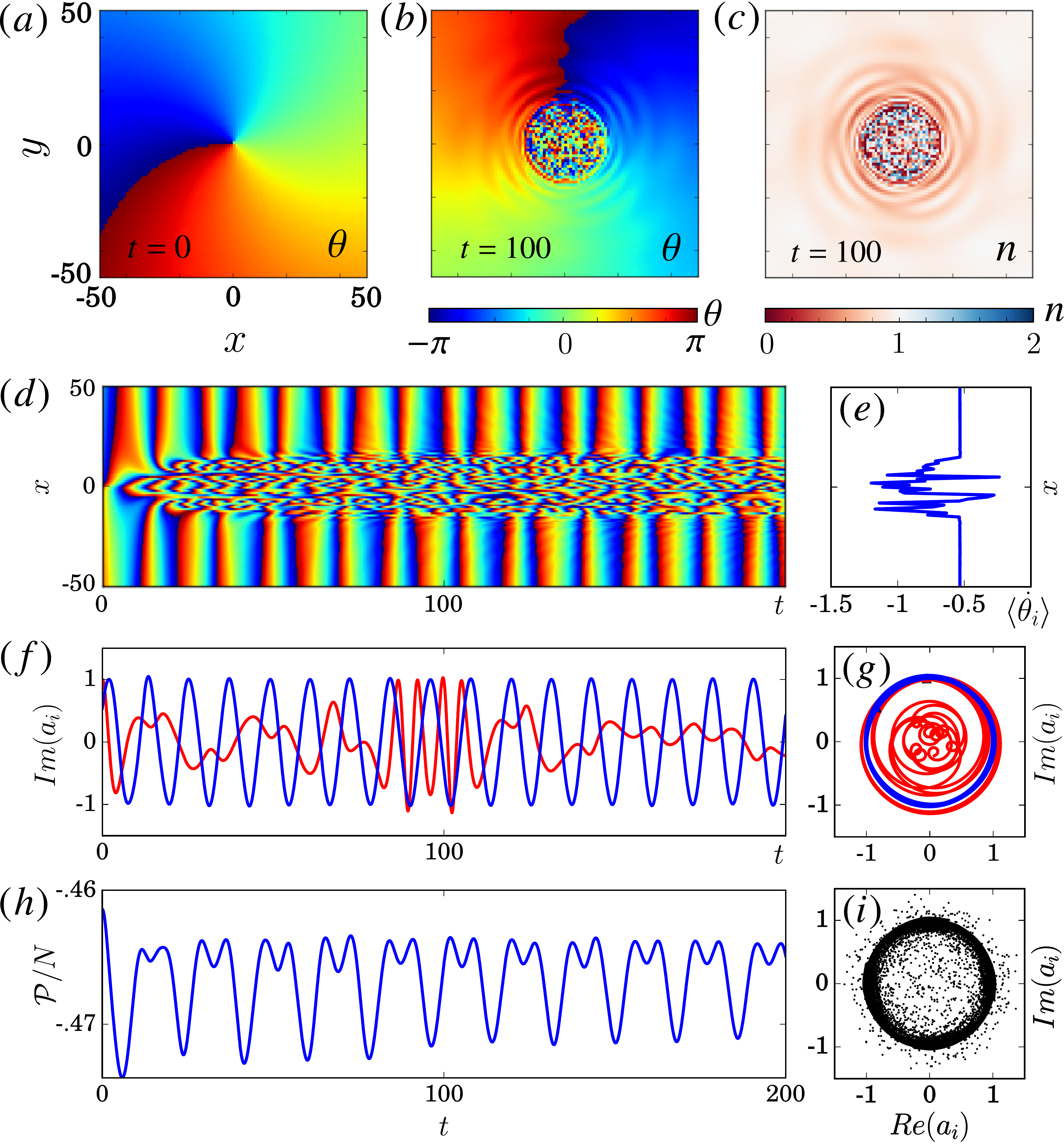}
\par\end{centering}

\centering{}\caption{\label{fig: nlhm-dynamics}
Chimera patterns in the 2D NLHM given by Eq.~\eqref{eq: NLHM}.
(a) The initial phase with uniform amplitude $|a_i|=1$ at time $t=0$.
(b,c) Phase $\theta_i$ and number of particle $n_i=|a_i|^{2}$ at $t=100$.
(d) Time evolution of the phase $\theta_i(t)$ for the cross-section $y=0$.
(e) Averaged local rotation speed $\langle\dot{\theta}_{i}\rangle$ over the time interval in (d).
(f) Time evolution of the points near the center $(x,y)=(-5,0)$ (red) and far away $(-100,0)$ (blue).
(g) Local phase space trajectory of (f).
(h) Hopping energy per particle $\mathcal{P}/\mathcal{N}$ variation over time.
(i) Phase portrait of all points at $t=100$.
Parameters: $Un_{0}=1$, $P=0.5$, $R=16$, and length $L=256$ with no-flux boundary condition.
Only the core region is shown for clarity. The hopping kernel $G_{ij}$ is given in Table \ref{tab: hopping-kernel}. Dimensionless units and $\hbar=1$ are used. 
}
\end{figure}

\begin{figure}
\begin{centering}
\includegraphics[width=0.9\columnwidth]{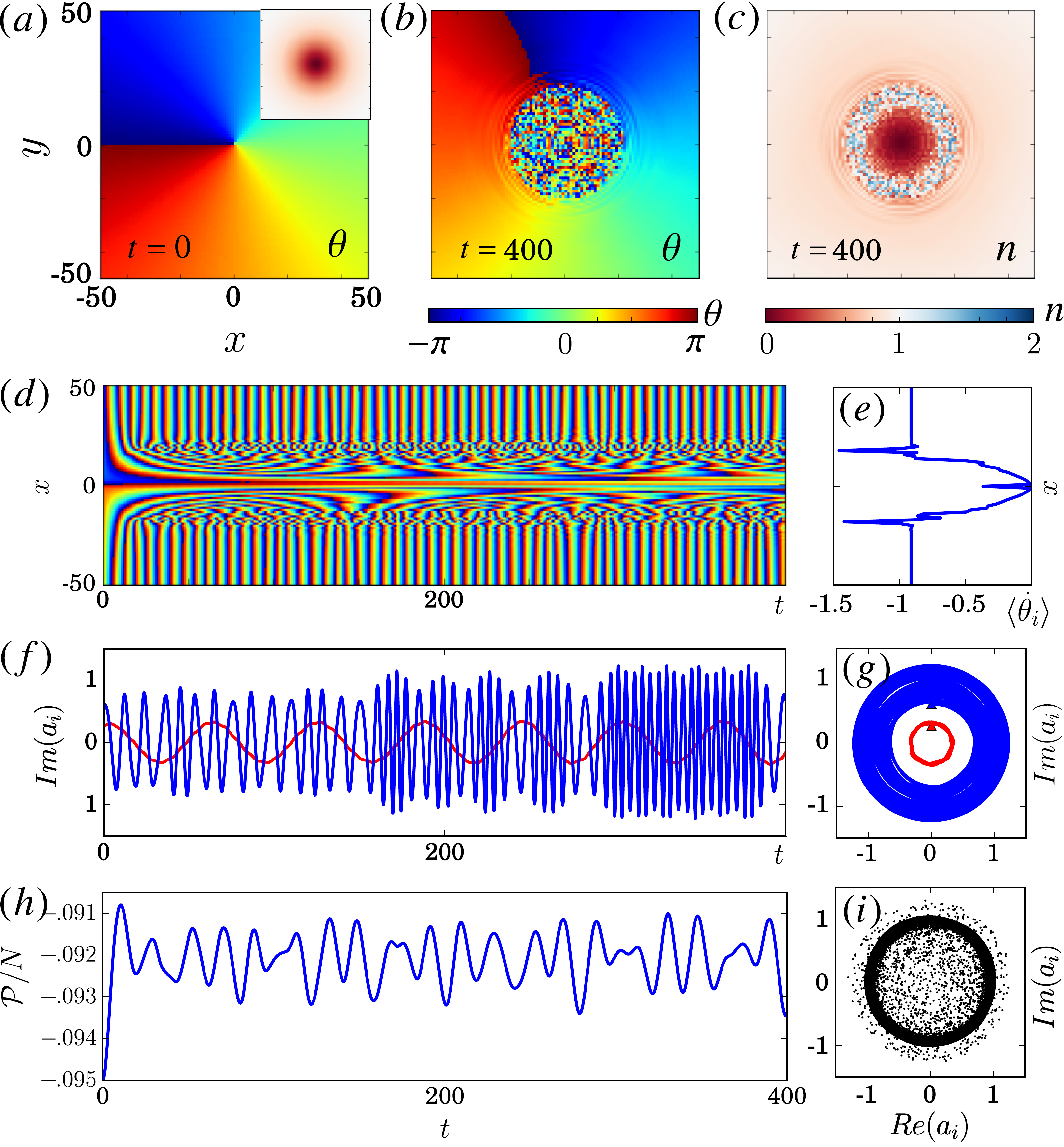}
\par\end{centering}

\centering{}\caption{\label{fig: nlhm-dynamics-vortex}
Similar to Fig. \ref{fig: nlhm-dynamics} but with vortex initial condition given by the phase in (a) and the density in the inset, and with a weaker hopping strength.
The points are $(x,y)=(-5,0)$ (red) and $(-15,0)$ (blue) in (f). 
For (g) and (i), $t=100$.
Parameters: $Un_{0}=1$, $P=0.1$, $R=16$, and size $L=256$ with no-flux boundary condition.
}
\end{figure}

Similarly to the 1D system, an initial condition with random phase regions can sustain itself over time in 2D.
Here, we focus on such chimera patterns, in particular those where an incoherent region forms spontaneously around a phase singularity \cite{martens_solvable_2010,gu_spiral_2013,lau_linked_2016}.
These patterns benefit from a topological protection in the sense that the incoherent core is robust against fluctuations in the phases.
The first initial condition we examine is a spiral phase initial condition that is locally phase coherent everywhere except the center, with uniform density, as illustrated in Fig.~\ref{fig: nlhm-dynamics}a (see also Methods). 
With this initial condition, the system can spontaneously evolve into a state with a small incoherent core surrounded by a large spatially coherent region as shown in Fig.~\ref{fig: nlhm-dynamics}b for the phase field.
Moreover, the density is randomized near the same core region in Fig.~\ref{fig: nlhm-dynamics}c.
As shown by the dynamics of a cross-section in Fig. \ref{fig: nlhm-dynamics}d, this spatial structure is sustained over long times (see Fig. S4 for snapshots and animations in SM). 
In addition, the same patterns can be observed even when the system size $L$ and also $R$ are increased (see Fig. S5 in SM).
The local dynamics of the two oscillators in Figs. \ref{fig: nlhm-dynamics}f and \ref{fig: nlhm-dynamics}g clearly shows the difference between two regions: $a_{i}$ oscillates regularly far from the core, but not close to it.
As in the 1D system, the incoherent region can only appear if the hopping range $R$ is sufficiently large, here $R \gtrsim 3$.
Moreover, with nearest-neighbor hopping, the system reduces to the discrete GPE so that the incoherent region spreads out and interferes like a wave (see Fig. S8 in SM).
All of these features are consistent with previous observations of chimera cores for driven-dissipative systems with self-sustained oscillators~\cite{kim_pattern_2004,martens_solvable_2010,lau_linked_2016}.
The distinct features in 2D are similar to the ones in 1D. This is shown in Fig. \ref{fig: nlhm-dynamics} (e)-(g) for the angular frequencies and the trajectories in phase space. Note especially the strong variations in the average local rotation speed. 
In particular, the oscillators can exhibit significant variations in amplitudes as follows from Fig. \ref{fig: nlhm-dynamics}(c),(g) and the phase portrait in Fig. \ref{fig: nlhm-dynamics}(i) that shows the phase and amplitude of all oscillators at a given moment in time.
We would like to point out that after the formation of the chimera core, the pattern persists over the longest time scales we were able to simulate ($>1000$ spiral rotations). This observation suggests that if a random phase core is used as an initial condition, the chimera core pattern also persists over such long times scale. This is indeed what we observe (see Fig. S6 in SM).

The important amplitude-dependent dynamics without limit cycles can be clearly observed for the vortex phase initial condition with amplitude going to zero at center in Fig. \ref{fig: nlhm-dynamics-vortex} (see Fig. S7 for snapshots in SM), with a weaker hopping $P=0.1$.
Similar to the 1D case discussed above, the fluctuations in the amplitude remain close to the initial condition for small $P$.
In particular, oscillators with different amplitudes have different oscillating frequencies even in the weak hopping regime according to Eq. (\ref{eq: phase-density-equation}) with small corrections arising from the weak hopping.
More importantly, as a conservative Hamiltonian system, it has time reversal symmetry and it conserves both quantities $\mathcal{H}$ and $N$ (see Fig. S9 and animations in SM).
This leads to persistent fluctuations or ripples as observed in Fig.~\ref{fig: nlhm-dynamics}b-d, which would be damped away in a dissipative system quickly.
In addition, the results of the backward time evolution of the core region are very delicate.
With a small perturbation, the background can evolve back to nearly the same state at $t=0$, but the core remains incoherent (see Fig. S9 in SM), which again signifies the difference between the two regions (see Sec. S4 and animations in SM).
This suggests that the Poincar\'e recurrence time to a regular spiral --- the time it takes to return within an arbitrarily small but finite distance to the original state (modulo possible rotations or translations) --- is large and that the probability to encounter a regular spiral is zero in the infinite system size limit.

Moreover, the hopping energy $\mathcal{P}$ is not constant as shown in Fig. \ref{fig: nlhm-dynamics}(h) even though the total energy $\mathcal{H}$ is constant.
Hence, there is a conversion between $\mathcal{P}$ and $\mathcal{U}$ over time.
This is different from a simple coherent and uniform distribution $a_i=\sqrt{n_0}$ having an energy per particle given by $\mathcal{H}/N = Un_0^2/2 - Pn_0$ with constant $\mathcal{P}$ and $\mathcal{U}$.
Note that all chimera patterns considered here do not correspond to ground states of the Hamiltonian but are excited states.

In realistic experimental systems, a small amount of particle loss typically exists and can be modeled phenomenonlogically by the term $U\to U-iU_{loss}$. 
Intuitively, the dynamics should not change significantly if the loss of the particles is less than half of the initial number of particles given by the condition $U_{loss} t/\hbar \lesssim 1$,
Indeed, chimera patterns can, for example, still be observed with $U_{loss}/U=0.02$ at a sufficiently short time (Fig. S11 in SM).
Further details about such loss in 1D (Fig. S10 in SM) and 2D (Fig. S11 in SM) are discussed in Sec. S4 in SM.

\subsection*{Mechanism for nonlocal hopping and the minimal model}

\subsubsection*{Mediating mechanism}

\begin{figure}
\begin{centering}
\includegraphics[width=0.9\columnwidth]{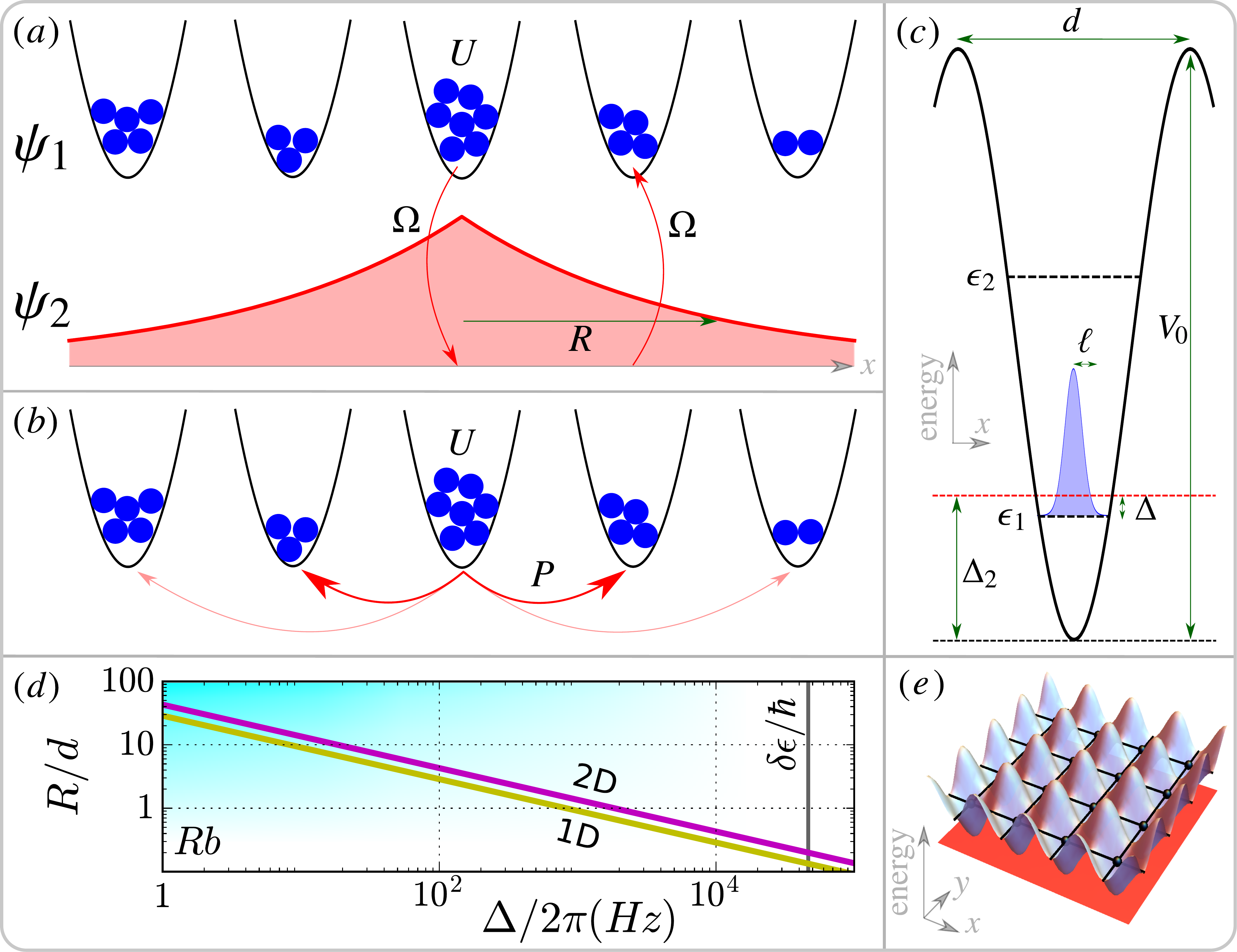}
\par\end{centering}

\caption{\label{fig: illu-nonlocal-hopping} Illustration of mediated hopping.
(a) Two-component model: Particles with on-site interaction $U$ are trapped (denoted by $\psi_1$) but can be converted into a mediating state (denoted by $\psi_2$) that can propagate freely.
It is eventually converted back to nearby sites, giving rise to a characteristic hopping range $R$.
(b) Effective model with hopping strength $P$ after adiabatically eliminating the fast mediating channel.
(c) Periodic lattice with spacing $d$ and lattice depth $V_{0}$: Trapped bosonic particles can be described by local ground state wavefunctions with width $\ell$ and energy $\epsilon_{1}$ (with energy gap $\delta\epsilon = \epsilon_{2}-\epsilon_{1}$).
$P$ and $R$ can be controlled by the Rabi frequency $\Omega$ and the detuning $\Delta=\Delta_{2}-\epsilon_{1}/\hbar$ between localized states and mediating states.
(d) $R$ can be adjusted by $\Delta$, see text for details.
(e) 2D periodic lattice considered in Fig. \ref{fig: bec-chimera}.
}
\end{figure}

The key idea for the mediating mechanism is to attach an inter-convertible mediating channel (labelled by $\psi_2$) to trapped states (labelled by $\psi_1$) as illustrated in Fig.~\ref{fig: illu-nonlocal-hopping}a.
With direct hopping, increasing the energy barrier between neighboring sites decreases both the hopping strength and the hopping range together.
In contrast, if the particles can be converted into fast mediating states that do not experience any energy barrier, then the particles can physically jump much further away.
Mathematically, this channel can be eliminated adiabatically (as done, for example, in~\cite{kuramoto_coexistence_2002,laing_chimeras_2015} for non-Hamiltonian systems), resulting in an effective nonlocal model (see Fig.~\ref{fig: illu-nonlocal-hopping}b) with independently adjustable on-site nonlinearity, hopping strength, and hopping range that can be tuned from nearest-neighbor to global hopping.

\subsubsection*{Minimal model}

A minimal mathematical model that captures the concepts of the mediating channel discussed above takes the form:
\begin{subequations}\label{eq: GPE-minimal}
\begin{eqnarray}
i\hbar\dot{\psi}_{1}(\mathbf{r},t) & = & U|\psi_{1}|^{2}\psi_{1}+\hbar\Omega\psi_{2}\label{eq: GPE-localized}\\
i\hbar\dot{\psi}_{2}(\mathbf{r},t) & = & -\hbar\kappa\nabla^{2}\psi_{2}+\hbar\Omega\psi_{1}+\hbar\Delta\psi_{2}\label{eq: GPE-mediating}
\end{eqnarray}
\end{subequations}
for the localized $\psi_1$ and mediating $\psi_2$ components respectively.
The corresponding Hamiltonian is given in Eq. (\ref{eq: generic-BEC}) with appropriate parameters. 
The inter-conversion is governed by a coherent coupling with Rabi frequency $\Omega$ and a detuning $\Delta$ that conserves the number of particles \cite{scully_quantum_1997,gerry_introductory_2005}.
Eq. (\ref{eq: GPE-mediating}) is essentially the Schr\"odinger equation for free particles with inverse mass $\kappa=\hbar/(2m)>0$ and so the particles can propagate outward.
The additional detuning in the far-detuned regime $|\Delta| \gg |\Omega|$ can ensure the mediating idea is well-defined:
The number of particles $N_{j}=\int d\mathbf{r}|\psi_{j}|^{2}$ in the mediating channel $N_{2}\ll N_{1}\approx N$ can be neglected.
Note that this model is not captured by the framework of nonlocal diffusive coupling \cite{tanaka_complex_2003}.
It is explicitly constructed to always preserve the conservation properties of the underlying Hamiltonian system, even when adiabatic elimination is applied.

\subsubsection*{Adiabatic elimination}

\begin{table}
\caption{\label{tab: hopping-kernel} The $D$-dimensional Green's function $G_{D}(r)$ with $r=|\mathbf{r}_{j}-\mathbf{r}_{i}|$.
$K_0$ is the modified Bessel function of the second kind.
}

\centering{}
\begin{tabular}{lccc}
\hline
 & \qquad{}$D=1$\qquad{} & \qquad{}$D=2$\qquad{} & \qquad{}$D=3$\qquad{}\quad{}\\
\hline
\hline
$G_D$ & $e^{-r/R}$ & $K_{0}(r/R)$ & $\frac{1}{r}e^{-r/R}$\\
\hline
\end{tabular}
\end{table}

\begin{figure}
\begin{centering}
\includegraphics[width=0.66\columnwidth]{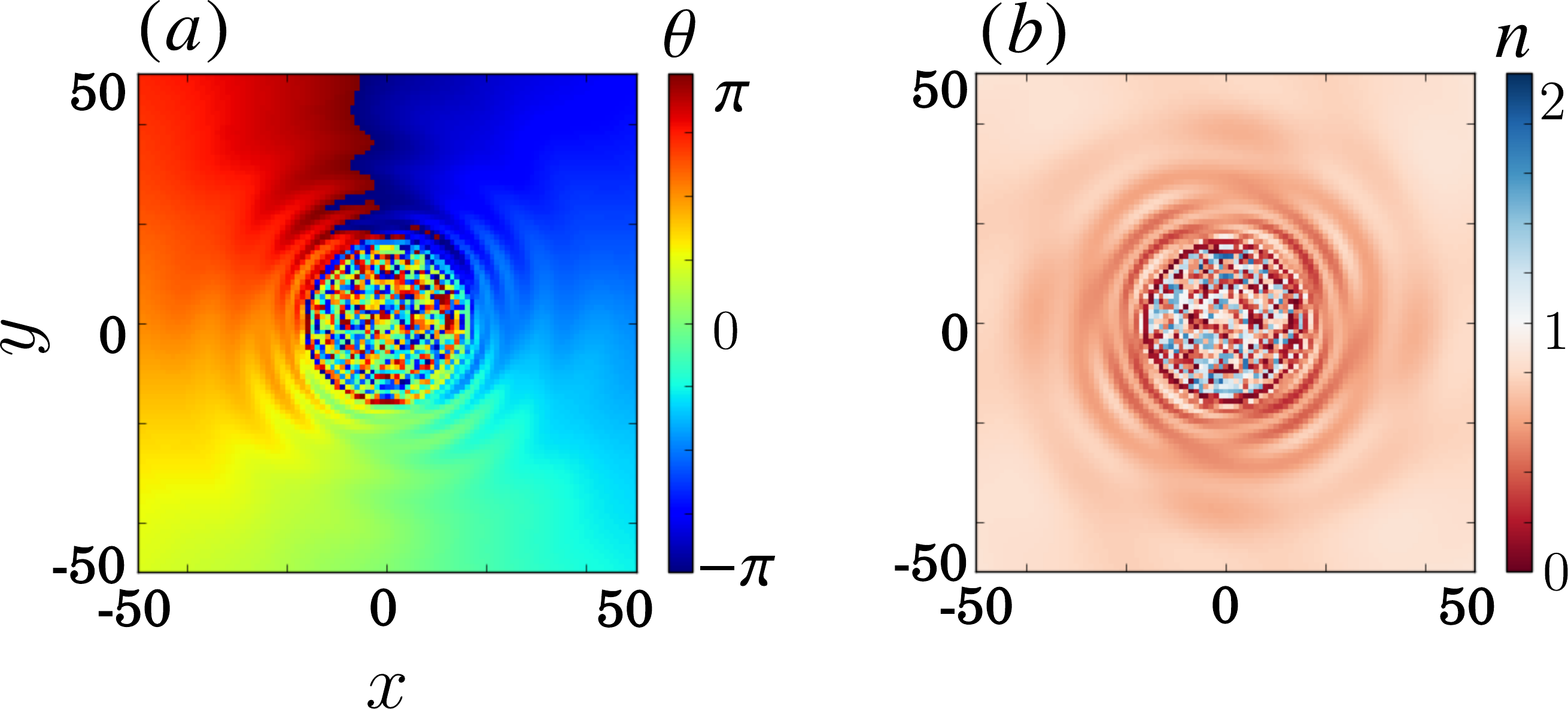}
\par\end{centering}

\centering{}\caption{\label{fig: mm-chimera}
Chimera patterns in the minimal model with the direct simulation using Eq.~\eqref{eq: GPE-minimal} at $t=100$ similar to Fig. \ref{fig: nlhm-dynamics}b-c.
The setting is the same as in Fig. \ref{fig: nlhm-dynamics} but with parameters $\Delta=16$, $\Omega=\sqrt{8}$, $U=1$, and $\kappa=4096$.
}
\end{figure}

Suppose $\psi_{1}$ evolves much slower than $\psi_{2}$, then we can apply adiabatic elimination by setting $\dot{\psi}_{2}=0$ \cite{brion_adiabatic_2007}.
The solution of $-\kappa\nabla^{2}\psi_{2}+\Omega\psi_{1}+\Delta\psi_{2}=0$ in the unbounded isotropic space with translation invariant is given by the convolution $\psi_{2}(\mathbf{r},t)=-(\Omega/\Delta)G_{D}(\mathbf{r})*\psi_{1}(\mathbf{r},t)$, where $G_{D}(\mathbf{r})$ is the $D$-dimensional hopping kernel as listed in Table \ref{tab: hopping-kernel}, with hopping radius $R=\sqrt{\kappa/|\Delta|}$.
Note that $\Delta>0$ is required for the solution of confined hopping kernels (see the form of $\psi_2$ in Fig. \ref{fig: illu-nonlocal-hopping}a), while $\Delta<0$ leads to wave-like solution.
Substituting this solution back into Eq. (\ref{eq: GPE-localized}), we can get the continuum NLHM:
\begin{equation}
i\hbar\dot{\psi}(\mathbf{r},t) = U|\psi|^2\psi - P\int d\mathbf{r}'G(\mathbf{r},\mathbf{r}')\psi(\mathbf{r}',t),
\end{equation}
where the summation is replaced by an integral with hopping strength $P=\hbar\Omega^{2}/\Delta$.
As shown in Fig. \ref{fig: mm-chimera}, the continuous NLHM well-approximates the discrete NLHM results from Fig. \ref{fig: nlhm-dynamics}.

\subsection*{Mediated hopping in ultracold atomic systems \label{sec:Ultracold-atomic-system}}

\subsubsection*{Hamiltonian and dynamic equation}

An ultracold atomic system of a general two-component GPE in a spin-dependent trap with coherent conversion is given by the Hamiltonian:
\begin{equation}
\mathcal{H}=\sum_{i=1,2}\left(\mathcal{H}_{i}+\frac{1}{2}\mathcal{U}{}_{ii}\right)+\mathcal{U}{}_{12}+\mathcal{R}, \label{eq: generic-BEC}
\end{equation}
with
\begin{eqnarray}
\mathcal{H}_{i} & = & \int d\mathbf{r}\left(\frac{\hbar^{2}}{2m_i}|\nabla\psi_{i}(\mathbf{r})|^{2}+V_{i}(\mathbf{r})|\psi_{i}(\mathbf{r})|^{2}\right),\\
\mathcal{U}_{ij} & = & g_{ij}\int d\mathbf{r}|\psi_{i}(\mathbf{r})|^{2}|\psi_{j}(\mathbf{r})|^{2},\\
\mathcal{R} & = & \sum_{i=1,2}\hbar\Delta_{i}\int d\mathbf{r}|\psi_{i}(\mathbf{r})|^{2}+\hbar\Omega\int d\mathbf{r}\left(\psi_{1}^{*}(\mathbf{r})\psi_{2}(\mathbf{r})+\psi_{2}^{*}(\mathbf{r})\psi_{1}(\mathbf{r})\right),
\end{eqnarray}
and with the normalization $N=N_{1}+N_{2}$ where $N_{i}=\int d\mathbf{r}|\psi_{i}(\mathbf{r})|^{2}$ is the number of particles for each component. 
$m_i$ is the mass of the particles, $V_{i}(\mathbf{r})$ is the trap potential, $g_{ij}$ is the two-particle collision coefficient, and we assume $g_{12}=g_{22}=0$ for the moment (see explanation below for non-zero case).
The Rabi oscillation term $\mbox{\ensuremath{\mathcal{R}}}$ represents the inter-conversion between the two components with the spatially homogeneous Rabi frequency $\Omega$ and the detuning $\Delta_{i}$.
By setting $V_i=0$, $m_1 \to \infty$, and $\Delta_{1}=0$, we arrive at the Hamiltonian for the minimal model discussed above.
When a small nonlinearity exists in the mediating channel, the effective detuning becomes $\Delta \to \Delta+g_{12}|\psi_1|^2+g_{22}|\psi_2|^2$ if $\psi_i$ is uniform.
Hence, the hopping radius decreases for $g_{ij}>0$ which is typical for atomic systems.
Note that when $|\psi_i|^2$ is small, the nonlinear effect can be ignored.
It can be achieved by decreasing the density, which is one of the main technique used in the analysis of real systems below.

Mathematically, Eq. (\ref{eq: GPE-localized}) can be obtained by setting appropriate parameters for the system described by Eq. (\ref{eq: generic-BEC}).
In particular, the absence of kinetic energy term in Eq. (\ref{eq: GPE-localized}) requires $m_1 \to \infty$.
However, the mass $m$ of inter-convertible atomic systems are the same, so $m_i = m$.
To circumvent this, we can increase the effective mass; for example, by placing the atoms in a periodic lattice.
This can be achieved by additionally setting $V_2=0$, $V_{1}$ to be periodic, and $\Delta_1=0$. Then the dynamic equation becomes \cite{nicklas_rabi_2011}:
\begin{subequations}\label{eq: GPE-trap}
\begin{eqnarray}
i\hbar\dot{\psi}_{1}(\mathbf{r},t) & = & \left(-\hbar\kappa\nabla^{2}+V_{1}+g_{11}|\psi_{1}|^{2}\right)\psi_{1}+\hbar\Omega\psi_{2}\label{eq: GPE-trap-localized}\\
i\hbar\dot{\psi}_{2}(\mathbf{r},t) & = & \left(-\hbar\kappa\nabla^{2}+\hbar\Delta_{2}\right)\psi_{2}+\hbar\Omega\psi_{1}\label{eq: GPE-trap-mediating}
\end{eqnarray}
\end{subequations}
Only the positive detuning $\Delta = \Delta_{2} - \epsilon_{1}/\hbar>0$ is considered here as illustrated in Fig. \ref{fig: illu-nonlocal-hopping}c.

\subsubsection*{Mapping to effective NLHM}

Note that direct adiabatic elimination does not work if states with high energy $\epsilon_{i>1}$ are occupied.
This is because high energy states do not evolve slowly compared to the mediating component.
To avoid occupying higher energy levels, we can confine the system to local ground states $\phi(\textbf{r})$ with energy $\epsilon_{1}$ and prevent excitation by choosing a suitable detuning such that $\epsilon_{2}-\epsilon_{1} \gg \hbar\Delta \gg \hbar|\Omega|$ (see Fig. \ref{fig: illu-nonlocal-hopping}c).
Under these  constraints, along with adiabatic elimination, we can show (Sec. S2 in SM) that Eqs. (\ref{eq: GPE-trap-localized}) and (\ref{eq: GPE-trap-mediating}) reduce to the exact form of Eq. (\ref{eq: NLHM-dynamic}) with $U=g_{11}\int |\phi|^4$, $P=\hbar\Omega^{2}/\Delta$, hopping kernel $G_{D}(r)$ in Table \ref{tab: hopping-kernel}, and
\begin{equation}
R=C_{D}\left(\frac{d}{2\ell}\right)^{\frac{D}{2}}\sqrt{\frac{\kappa}{\Delta}}\label{eq: R_eff_hopping_radius}
\end{equation}
for $d\gg2\ell$, where $C_{D}$ is a constant.
Intuitively, particles staying in the mediating channel for a longer time have a larger hopping range $R\sim\Delta^{-1/2}$.
Since the effective conversion region has a characteristic length scale $2\ell$ in a unit lattice with length $d$, scaling with $2\ell/d$ is expected.
Indeed, we have the effective scaling $\Delta\to\Delta_{eff}=(2\ell/d)^{D}\Delta$.
The self-consistency condition for adiabatic elimination is $\hbar\Delta\gg Un_{0},P$ assuming all $n_{i}\sim n_{0}$ ($n_{0}$ is the average number of particles per site).
In this effective NLHM, $a_{i}$ in Eq. (\ref{eq: NLHM}) represents the state of a localized wavepacket at site $i$.
Moreover, the kernel $G_{ij}$ in Eq. (\ref{eq: NLHM}) describes the matter-wave mediated hopping with wavepackets annihilated at site $j$ and created at site $i$.

\subsubsection*{Optical lattice \label{sec:Optical-lattice}}

The system discussed above requires a particle that is inter-convertible, which can be an atom with two different hyperfine states.
A candidate is the Rubidium atom with hyperfine states $|F=1, m_F=-1\rangle$ and $|F=1, m_F=0\rangle$ which has been realized in a spin-dependent trap \cite{mckay_slow_2013}.
Suppose the trapping potential is sinusoidal $V_{1}(\textbf{r})=V_{0}\sum_{\sigma}\sin^{2}(kx_{\sigma})$ with wavelength $\lambda$, wavenumber $k=2\pi/\lambda$, lattice spacing $d=\lambda/2$, and trap depth $V_{0}$.
The summation is taken over the lattice trap dimension as shown in Fig. \ref{fig: illu-nonlocal-hopping}c or \ref{fig: illu-nonlocal-hopping}e.
For sufficiently large $V_{0}$, all direct hopping can be suppressed, and the local ground states at trap minima can be approximated by a Gaussian $\phi_{\sigma}(x_{\sigma})=e^{-\pi x^{2}/(2\ell_{\sigma}^{2})}/\sqrt{\ell_{\sigma}}$ with $\ell_{\sigma}=\sqrt{\pi\hbar/(m\omega_{\sigma})}$.
In this setting, the nonlinearity is enhanced by the high density since  $U=g_{11}/W$ with effective volume $W=2^{3/2}\ell_{x}\ell_{y}\ell_{z}$.
The constant can be found by numerical fitting, which gives $C_{D}\approx1$ (see Sec. S3 and Fig. S1 in SM).

\subsubsection*{Achievable hopping range \label{sec:Hopping-range}}

For the hopping to be considered nonlocal, $R>d$ must be satisfied.
An example of Rubidium atoms is shown in Fig. \ref{fig: illu-nonlocal-hopping}d with $d=395$nm and a deep trap $s=40$ (expressing $V_{0}=sE_{R}$ in recoil energy $E_{R}=\hbar\kappa k^{2}$).
With such a large $s$, as studied before \cite{jaksch_cold_1998}, the overlap between wavefuncion of neighboring cell is very small,  the direct hopping is weak, and the system becomes a Mott insulator in the quantum regime.
Nevertheless, mediated hopping can completely replace the direct hopping (with order $R\sim d$, see Fig. \ref{fig: illu-nonlocal-hopping}d) and allow real time control.
Since $\Omega$, $\Delta$, and $U$ can be easily adjusted in experiments, there seems to be no upper bound on $R$.
From a practical point of view, however, it is limited by the lifetime $\tau$ and experimental duration.
A simple estimation of $\tau\sim 1$s gives a maximum $R\sim 30d$ as shown in Fig. \ref{fig: illu-nonlocal-hopping}d.

\subsubsection*{Tuning nonlinearity and loss \label{sec:Nonlinearity-and-loss}}

The regime with competitive $P\sim Un_{0}$ is the most interesting.
However, a BEC in a 3D optical lattice using the parameters given above has a strong nonlinearity $U/\hbar=2\pi\times2.23\mbox{kHz}$, which demands a large $\Delta$ and, consequently, a small $R$.
$U$ can be reduced by the use of two tuning techniques: Decreasing the density, or utilizing the Feshbach resonance.
The latter method can experimentally tune the nonlinearity over many orders of magnitude \cite{chin_feshbach_2010}.
The former method is preferable because both nonlinearity and collision loss can be decreased simultaneously.
In 1D and 2D lattices, the non-lattice dimension can be weakly trapped to reduce the density, resulting in a lattice of disk and cigarette-shaped wavefunctions respectively \cite{burger_dark_1999,bloch_ultracold_2005}.
In this case, the dominant loss is the two-particle loss in the localized component.
The rate of the two-particle loss can be estimated by $U_{loss}=\hbar L_{11}/W$ and therefore half-life $\tau=W/L_{11}$ with two-particle loss rate $L_{11}$ \cite{lau_proposal_2014}. This implies that $\tau\sim\ell_{z}$ in 2D, so increasing $\ell_{z}$ can improve the BEC lifetime.

\subsubsection*{Chimera patterns in BECs \label{sec:Chimera-states-in-BECs}}

\begin{figure}
\begin{centering}
\includegraphics[width=0.9\columnwidth]{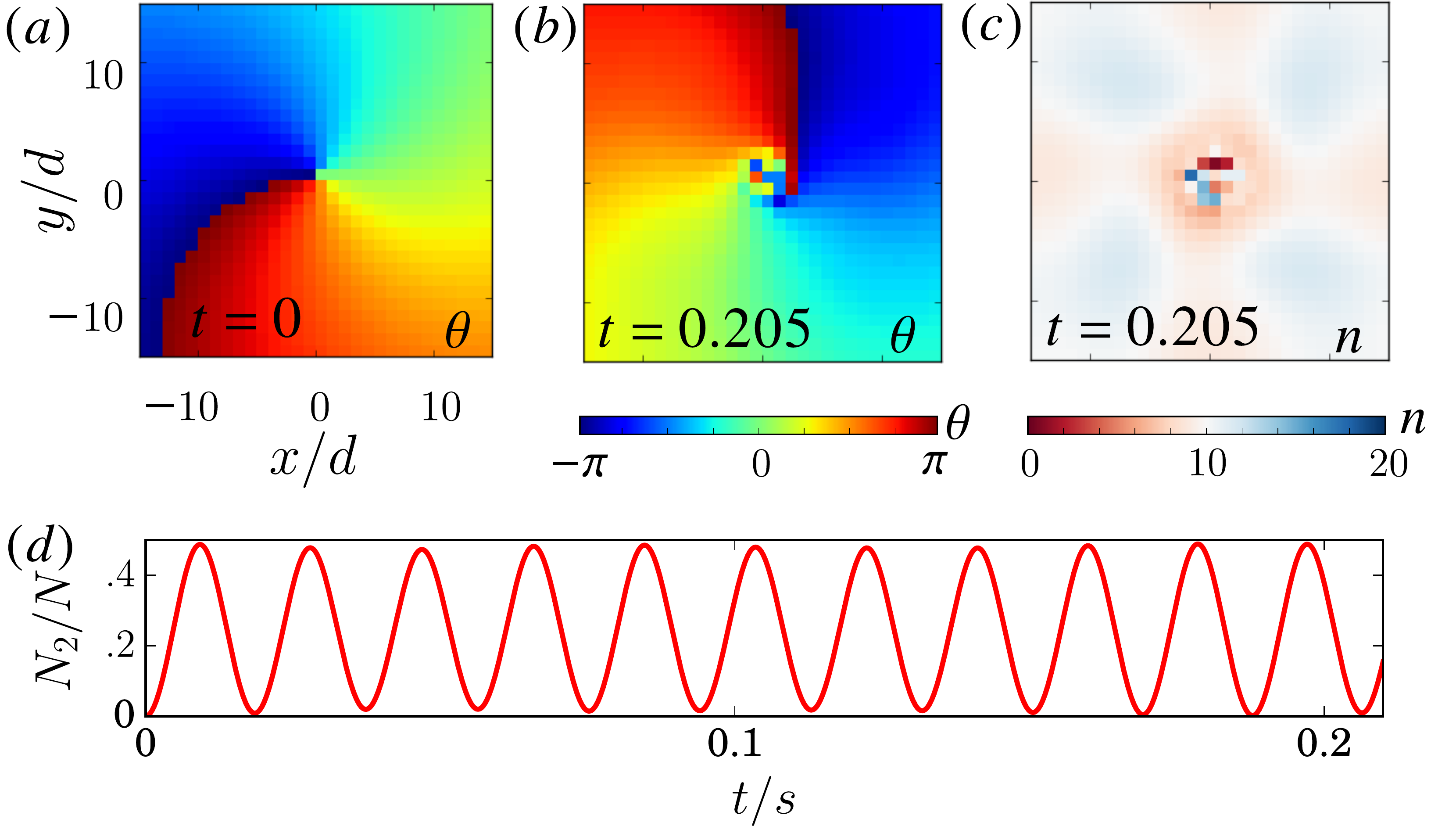}
\par\end{centering}

\centering{}\caption{\label{fig: bec-chimera}
Chimera patterns in BECs. 
(a) Initial phase $\theta_i$ with a uniform number of particles per lattice site $n_i=10$.
(b,c) $\theta_i$ and $n_i$ of the state at time $t=205$ms.
The simulation is based on Eq. (\ref{eq: GPE-trap}) in the 2D lattice given by Fig. \ref{fig: illu-nonlocal-hopping}e with $100d\times 100d$ and the no-flux boundary condition.
We spatially average over the lattice units.
(d) Inter-conversion between two components.
For the optical lattice, we use Rubidium $^{87}$Rb with $s=40$ and $d=395\mbox{nm}$ which gives and $\ell_{x}=\ell_{y}=0.22d$.
Additional parameters: $\Delta=2\pi\times48\mbox{Hz}$, $\Omega=2\pi\times32\mbox{Hz}$. The density is decreased by using $\ell_{z}=200\ell_{x}$, and the nonlinearity is weaken 10 times by using Feshbach resonance.
The estimated values are $Un_{0}/\hbar\approx2\pi\times19\mbox{Hz}$, $P\approx2\pi\times16\mbox{Hz}$, $R\approx6d$, and $\tau\approx 5\mbox{s}$.
}
\end{figure}

The derivation of effective models implies that chimera patterns can also be observed in certain parameter regimes for Eqs. (\ref{eq: GPE-minimal}) and (\ref{eq: GPE-trap}).
The question is: can such parameter regimes be achieved experimentally?
The possible existence of chimera patterns in ultracold atoms is established in a parameter regime given in Fig. \ref{fig: bec-chimera}, based on a full simulation of Eqs. (\ref{eq: GPE-trap}).
Similar to Fig. \ref{fig: nlhm-dynamics}, a random core appears eventually.
Fig. \ref{fig: bec-chimera}d shows the Rabi oscillation between the two components with frequency $\sim\sqrt{\Omega^2+\Delta^2}$.
Note that most of the atoms can be converted back after a full period, which confirms the physical picture discussed in Fig. \ref{fig: illu-nonlocal-hopping}a and is consistent with previous works \cite{de_vega_matter-wave_2008,navarrete-benlloch_simulating_2011}.
We also observe chimera patterns in both the far-detuned regime and the regime with $\Delta\sim\Omega$, which may not be well-described by an effective NLHM and will be reported in a separate paper. 

Experimentally, the initial state can be prepared starting from a uniform BEC. Thousands of optical lattice sites \cite{bloch_ultracold_2005,sherson_single-atom-resolved_2010,wurtz_experimental_2009} can be created with $V_{1}$ adiabatically turned on until the direct hopping is suppressed and mediated hopping begins to dominate.
The energy shift induced by a short light-pulse can then be used to create any desired initial phase.
The system states and dynamics may be detected by using various techniques such as optical readout, time of flight techniques, or matter-wave interference \cite{davis_bose-einstein_1995,gadway_probing_2012}.
The loss $U_{loss}/U \approx 0.017$ here is comparable with the discussion in the minimal model.
Note that a small amount of loss can cause the BEC system to follow the classical trajectory \cite{habib_decoherence_1998}, and so each site can be well-described by a classical mean-field amplitude and phase.
At the same time, our simulations suggest that chimera core patterns in 2D are particularly robust due to their topological structure. Specifically, if we start with a chimera core initial condition it can persist over long times. This is particularly useful if the lifetime of BECs is further limited in a given experiment by other experimental imperfections.
All of this suggests that chimera patterns should be observable in experimental BECs.

\section*{Discussion}

In summary, we have shown the formation of chimera patterns in three conservative Hamiltonian systems.
Our work can be considered as a Hamiltonian formulation of chimera states, following a similar attempt for synchronization \cite{witthaut_kuramoto_2014}.
The NLHM is a direct analogue of the nonlocal CGLE \cite{kuramoto_coexistence_2002}, in that both systems have a third-order nonlinearity, except that the nonlocal diffusive coupling in the CGLE is replaced by nonlocal hopping in the NLHM.
The nonlocal hopping requires the underlying physical model to conserve both the energy and particle number, as it is typical in ultracold systems.
The results of our simulations with realistic parameters suggest that chimera patterns should be observable in the experiments of ultracold systems.
With only local hopping, the incoherent region will smooth out over time.
Therefore, the persistence of the incoherent region, and the formation of a chimera core starting from a vortex or spiral initial condition in 2D, are two distinct signatures of the correct implementation of the mediated nonlocal hopping.

Our results in this paper are based on classical conservative Hamiltonian systems, which provides a new avenue to understand chimera patterns.
These results may be extended into the quantum regime, since all of the physical processes that we analyzed are coherent and conserve both energy and particles.
Eq. (\ref{eq: NLHM})-(\ref{eq: NLHM-dynamic}), (\ref{eq: GPE-minimal})-(\ref{eq: GPE-trap}) can be quantized, and Eq. (\ref{eq: NLHM}) becomes the Bose-Hubbard model with tunable mediated hopping \cite{de_vega_matter-wave_2008}.
This opens the door for the exploration of exotic condensed-matter states, such as supersolid states and quantum vortices with topological defect, in addition to other long-range effects  \cite{landig_quantum_2016,henkel_three-dimensional_2010}.
The technique that we presented suggests that experimental studies of the synchronization and chimera patterns of a large number of oscillators may be feasible in quantum systems \cite{mari_measures_2013,lee_quantum_2013,walter_quantum_2014,bastidas_quantum_2015,witthaut_classical_2017,viennot_quantum_2016}. 
We hope that our work here motivates further studies on chimera states and mediated nonlocal hopping.

\section*{Methods}

The numerical methods we used are the fourth-order time splitting method for Gross-Pitaevskii equations \cite{antoine_computational_2013}.
This method for the conservative systems automatically conserve the particle number.
For systems with particle loss, we used the standard fourth-order Runge-Kutta is used.
The geometry used in the simulations is a square lattice with size $L$ and the no-flux boundary condition.
For the spiral initial condition, uniform density $|a_{i}|=\sqrt{n_{0}}$ is used and the state is given by $a_{i}(t=0)=\sqrt{n_{0}}e^{i(k_{s}r-\tan^{-1}(y/x))}$ with $r=\sqrt{(x^{2}+y^{2})}$.
For the vortex-like initial condition, the state is given by $a_i(t=0) = A_i e^{i \tan^{-1}(y/x)}$ with $A_i = 1-e^{-r/R_{vortex}}$ and $R_{vortex}$ is the length scale of the vortex.
We use $R_{vortex}=R$ in this manuscript.
For the system with a mediating channel, the channel is initially empty $\psi_2=0$.

\section*{Acknowledgements}

We thank David Hobill, Lindsay Leblanc, Matthew Fisher, Stephen Wein, Farokh Mivehvar, for useful discussions.
This research was enabled in part by support provided by WestGrid, Calcul Qu\'ebec, and Compute Canada.
H.W.H.L. was supported by AITF and NSERC.
J.D. and C.S. acknowledge financial support from NSERC.

\section*{Author contributions statement}

H.W.H.L. conceived the project, constructed the mathematical models, performed calculations and simulations, and prepared all figures and movies;
J.D. supervised the chimera pattern aspects, and wrote the chimera part of the main manuscript with H.W.H.L.;
C.S. supervised the atomic physics aspects, and wrote the atomic part of the main manuscript with H.W.H.L.;
All authors reviewed the final version of the manuscript.

\section*{Additional information}

Competing financial interests: The authors declare no competing financial interests.

\pagebreak
\clearpage
\widetext
\begin{center}
\textbf{\large Supplementary Materials: Chimera patterns in conservative systems and ultracold atoms with mediated nonlocal hopping}
\end{center}
\setcounter{equation}{0}
\setcounter{figure}{0}
\setcounter{table}{0}
\setcounter{page}{1}
\makeatletter
\renewcommand{\theequation}{S\arabic{equation}}
\renewcommand{\thefigure}{S\arabic{figure}}
\renewcommand{\bibnumfmt}[1]{[S#1]}
\renewcommand{\citenumfont}[1]{S#1}

\section{Alternative Hamiltonians of the nonlocal hopping model}

The Hamiltonian of the NLHM is given by 
\begin{equation}
\mathcal{H}=\frac{U}{2}\sum_{i}|a_{i}|^{4}-P\sum_{i,j}G_{ji}a_{i}^{*}a_{j},\label{eq: NLHM-supp}
\end{equation}
 with $G_{ij}=G_{ji}$ and $G_{ii}=0$. This Hamiltonian can be represented
in a few different canonical variables (see \cite{S_witthaut_kuramoto_2014,S_thommen_classical_2003}
for example). Suppose the canonical coordinate and momentum variables
are $q_{i}$ and $p_{i}$ respectively, then we can define
\begin{eqnarray}
a_{i} & = & \frac{1}{\sqrt{2}}(q_{i}+ip_{i}),\\
a_{i}^{*} & = & \frac{1}{\sqrt{2}}(q_{i}-ip_{i}).
\end{eqnarray}
With this transformation, the Hamiltonian becomes
\begin{equation}
\mathcal{H}=\frac{U}{8}\sum_{i}\left(q_{i}^{2}+p_{i}^{2}\right)^{2}-\frac{1}{2}P\sum_{i,j}G_{i,j}(q_{i}q_{j}+p_{i}p_{j}),
\end{equation}
Similarly, we can define the action and angle to be $n_{i}$ and $\theta_{i}$
respectively, such that $a_{i}=\sqrt{n_{i}}e^{i\theta_{i}}$, or
\begin{eqnarray}
n_{i} & = & \frac{1}{2}\left(q_{i}^{2}+p_{i}^{2}\right),\\
\theta_{i} & = & \tan^{-1}\left(p_{i}/q_{i}\right).
\end{eqnarray}
Now, the Hamiltonian becomes 
\begin{equation}
\mathcal{H}=\frac{U}{2}\sum_{i}n_{i}^{2}-P\sum_{i,j}G_{i,j}\sqrt{n_{i}n_{j}}\cos(\theta_{j}-\theta_{i}).
\end{equation}
Note that $n_{i}$ may be interpreted as the (mean-field) number of
particle at site $i$. Hence, the conservation of the total number
of particles implies the quantities $\sum_{i}|a_{i}|^{2}$, $\sum_{i}\left(q_{i}^{2}+p_{i}^{2}\right)$,
and $\sum_{i}n_{i}$ are constant. Moreover, the Hamiltonian is invariant
under the transformation $a_{i}\to a_{i}e^{i\theta_{0}}$ with arbitrary
global phase.

In the continuum limit, such as the adiabatic elimination of the simplest
two-component model in the the main text, the corresponding Hamiltonian
can be obtained by replacing $a_{i}\to\psi(\mathbf{r})$, $\sum_{i}\to\int d\mathbf{r}$,
$\sum_{i,j}\to\int\int d\mathbf{r}d\mathbf{r}'$ and $G_{i,j}\to G(\mathbf{r},\mathbf{r}')$.
Explicitly, the Hamiltonians are:
\begin{eqnarray}
\mathcal{H} & = & \frac{U}{2}\int d\mathbf{r}|\psi(\mathbf{r})|^{4}-P\int\int d\mathbf{r}d\mathbf{r}'G(\mathbf{r},\mathbf{r}')\psi^{*}(\mathbf{r})\psi(\mathbf{r}'),\\
\mathcal{H} & = & \frac{U}{8}\int d\mathbf{r}\left(q(\mathbf{r})^{2}+p(\mathbf{r})^{2}\right)^{2}-\frac{1}{2}P\int\int d\mathbf{r}d\mathbf{r}'G(\mathbf{r},\mathbf{r}')(q(\mathbf{r})q(\mathbf{r}')+p(\mathbf{r})p(\mathbf{r}')),\\
\mathcal{H} & = & \frac{U}{2}\int d\mathbf{r}\left(n(\mathbf{r})\right)^{2}-P\int\int d\mathbf{r}d\mathbf{r}'G(\mathbf{r},\mathbf{r}')\sqrt{n(\mathbf{r})n(\mathbf{r}')}\cos(\theta(\mathbf{r}')-\theta(\mathbf{r})).
\end{eqnarray}

\section{Hopping in ultracold atoms with a periodic lattice}

We start from the equations in the main paper: 
\begin{eqnarray}
i\hbar\dot{\psi}_{1}(\mathbf{r},t) & = & \left(-\hbar\kappa\nabla^{2}+V_{1}+g_{11}|\psi_{1}|^{2}\right)\psi_{1}+\hbar\Omega\psi_{2},\label{eq: GPE-trap-localized-SM}\\
i\hbar\dot{\psi}_{2}(\mathbf{r},t) & = & \left(-\hbar\kappa\nabla^{2}+\hbar\Delta_{2}\right)\psi_{2}+\hbar\Omega\psi_{1}.\label{eq: GPE-trap-mediating-SM}
\end{eqnarray}
In general, adiabatic elimination works best when the first component
evolves the slowest \cite{S_brion_adiabatic_2007}. However, no such
choice exist for an arbitrary wavefunction of Eq. (\ref{eq: GPE-trap-localized-SM}),
but it exists when the dynamics are confined to the lowest energy
band since the excitations have fast dynamics. We derive the effective
model with these two assumptions.

Note that there is no basis that is simultaneously good for both equations;
although, the good basis for the localized and mediating equation
are the Wannier basis and Fourier basis respectively. For the system
here, it is easier to understand in the Wannier basis $\{w_{mn}(\mathbf{r})\}$
\cite{S_jaksch_cold_1998,S_morsch_dynamics_2006} for a periodic lattice,
where $n$ is the energy band index and $m$ is the lattice site index.
In this new basis, the wavefunctions can be represented by $\psi_{1}(\mathbf{r},t)=\sum_{mn}a_{mn}(t)w_{mn}(\mathbf{r})$
and $\psi_{2}(\mathbf{r},t)=\sum_{mn}b_{mn}(t)w_{mn}(\mathbf{r})$
respectively. Substituting the transformation back into Eq. (\ref{eq: GPE-trap-localized-SM})
and (\ref{eq: GPE-trap-mediating-SM}), we have 
\begin{eqnarray}
i\hbar\dot{a}_{mn}(t) & = & \epsilon_{mn}a_{mn}+U|a_{mn}|^{2}a_{mn}+\hbar\Omega b_{mn},\qquad\mbox{for }n=1,\\
i\hbar\dot{b}_{mn}(t) & = & \hbar\sum_{kl}c_{mnkl}b_{kl}+\hbar\Delta_{2}b_{mn}+\hbar\Omega a_{mn},
\end{eqnarray}
where
\begin{eqnarray}
\epsilon_{mn} & = & \int_{V}d\mathbf{r}\left(\hbar\kappa|\nabla w_{mn}|^{2}+V_{1}|w_{mn}|^{2}\right),\\
U & = & g_{11}\int_{V}d\mathbf{r}|w_{mn}|^{4},\\
c_{mnkl} & = & \kappa\int_{V}d\mathbf{r}\nabla w_{mn}^{*}(\mathbf{r})\nabla w_{kl}(\mathbf{r}),
\end{eqnarray}
Note that we assume $a_{mn}=0$ for all $n>1$. We also assume that
the trap potential $V_{1}$ is sufficiently deep so that there is
no direct hopping. In this setting, the eigenenergy $\epsilon_{m1}=\epsilon_{0}$
is a constant. Hence, we can shift the energy $\Delta_{2}\to\Delta:=\Delta_{2}-\epsilon_{0}/\hbar$
using the transformation $a_{mn}\to a_{mn}e^{-i\epsilon_{0}t}$. If
the energy gap is large $\epsilon_{m2}-\epsilon_{m1}\gg\hbar\Delta$,
then we can ignore the resonance with the higher band index $n>1$.
Furthermore, with initially empty excited states, i.e. $a_{mn}(t=0)=0$
for $n>1$, no excited states will be populated because there are
no resonance with those states. Written explicitly:
\begin{eqnarray}
i\hbar\dot{a}_{m1}(t) & = & U|a_{m1}|^{2}a_{m1}+\hbar\Omega b_{m1},\\
i\hbar\dot{b}_{m1}(t) & = & \hbar\sum_{kl}c_{m1kl}b_{kl}+\hbar\Delta b_{m1}+\hbar\Omega a_{m1},\\
i\hbar\dot{b}_{mn}(t) & = & \hbar\sum_{kl}c_{mnkl}b_{kl},\qquad\mbox{for }n>1.
\end{eqnarray}
In this form, all the important dynamics are captured, and the localized
component can be slow relative to the mediating component.

\section{Hopping kernels with a lattice}

Suppose the mediating channel has a much faster time scale, so the
adiabatic elimination is the same as setting $\dot{b}_{mn}=0$. Therefore,
the hopping kernel can be found by solving $b_{m1}$ in the following
self-consistently equation by having $a_{m1}=1$ at the center:
\begin{eqnarray}
0 & = & \sum_{kl}c_{m1kl}b_{kl}+\Omega a_{m1}+\Delta b_{m1},\label{eq: discrete-screened-poisson-eq}\\
0 & = & \sum_{kl}c_{mnkl}b_{kl}\qquad\mbox{for }n>1,\nonumber 
\end{eqnarray}
This is the discrete analogue of finding the continuous hopping kernel
$G_{D}(\mathbf{r})$ as described in the main text by setting $\psi_{1}(\mathbf{r})=\delta(\mathbf{r})$,
except the interconversion only happens in certain regions. The effective
conversion regions have a length scale $2\ell$ of the localized wavefunction,
in each lattice unit with length $d$. Therefore, it is expected that
the solution $G_{ij}$ takes a similar form as the continuous system
with an effective scaling $\Delta\to\Delta_{eff}=(2\ell/d)^{D}\Delta$.
Hence, the solution is $b_{i1}=\frac{\Omega}{\Delta}G_{ij}*a_{j1}$.
Substituting back into the first component, the hopping strength becomes
\begin{equation}
P=\hbar\frac{\Omega^{2}}{\Delta},\label{eq: P_hopping_strength}
\end{equation}
the same as the continuous system, and $G_{ij}$ takes the same form
as in the Table 1 in the main text. The characteristic hopping radius
is 
\begin{equation}
R=C_{D}\left(\frac{d}{2\ell}\right)^{\frac{D}{2}}\sqrt{\frac{\kappa}{\Delta}},\label{eq: R_eff_hopping_radius}
\end{equation}
where $D$ is the dimension and $C_{D}$ is a constant.

The results above can be verified numerically. This requires a method
to find the hopping kernel in a periodic lattice self-consistently.
Here, we solve the corresponding time dependent equation of Eq. (\ref{eq: discrete-screened-poisson-eq})
and the solution is given by the equilibrium state. Hence, Eq. (\ref{eq: discrete-screened-poisson-eq})
with the time splitting method becomes  
\begin{eqnarray}
\dot{b}_{m1}(t) & =- & \Omega a_{m1}-\Delta b_{m1},\\
\dot{\tilde{\psi}}_{2}(\mathbf{q},t) & =- & \frac{\hbar q^{2}}{2m}\tilde{\psi}_{2},
\end{eqnarray}
for the conversion step and propagation step respectively, so the
basis is changed between each step. $\tilde{\psi}_{2}(\mathbf{q},t)$
is the wavefunction in Fourier space. The hopping kernel $G_{ij}$
is the same as the equilibrium solution $b_{i1}^{*}$ with $G_{mj}\sim b_{m1}^{*}$,
if the system is set to $a_{m1}=\delta_{j1}$, where $j$ is the source
lattice site (chosen to be the center of the lattice). For simplicity,
Gaussian approximation is used to approximate the lowest band Wannier
function as
\begin{equation}
w_{m1}(\mathbf{r})=\phi(\mathbf{r}-\mathbf{r}_{m})\sim e^{-\frac{\pi|\mathbf{r}-\mathbf{r}_{m}|^{2}}{2\ell^{2}}},
\end{equation}
which is a good approximation when $\ell\ll d$, such as deep sinusoidal
trap, where $\mathbf{r}_{m}$ is the center of the Gaussian. The transformation
between real space and Wannier basis are given by
\begin{equation}
b_{m1}=\langle w_{m1}(\mathbf{r})|\psi_{2}(\mathbf{r})\rangle=\int_{V}d^{3}\mathbf{r}\phi(\mathbf{r})\psi_{2}(\mathbf{r}),
\end{equation}
where $V$ is the lattice volume around the lattice minimum $[-d/2,d/2]^{D}$
with finite cutoff of lattice spacing $d$. 

\begin{figure*}
\begin{centering}
(a)\includegraphics[width=0.35\textwidth]{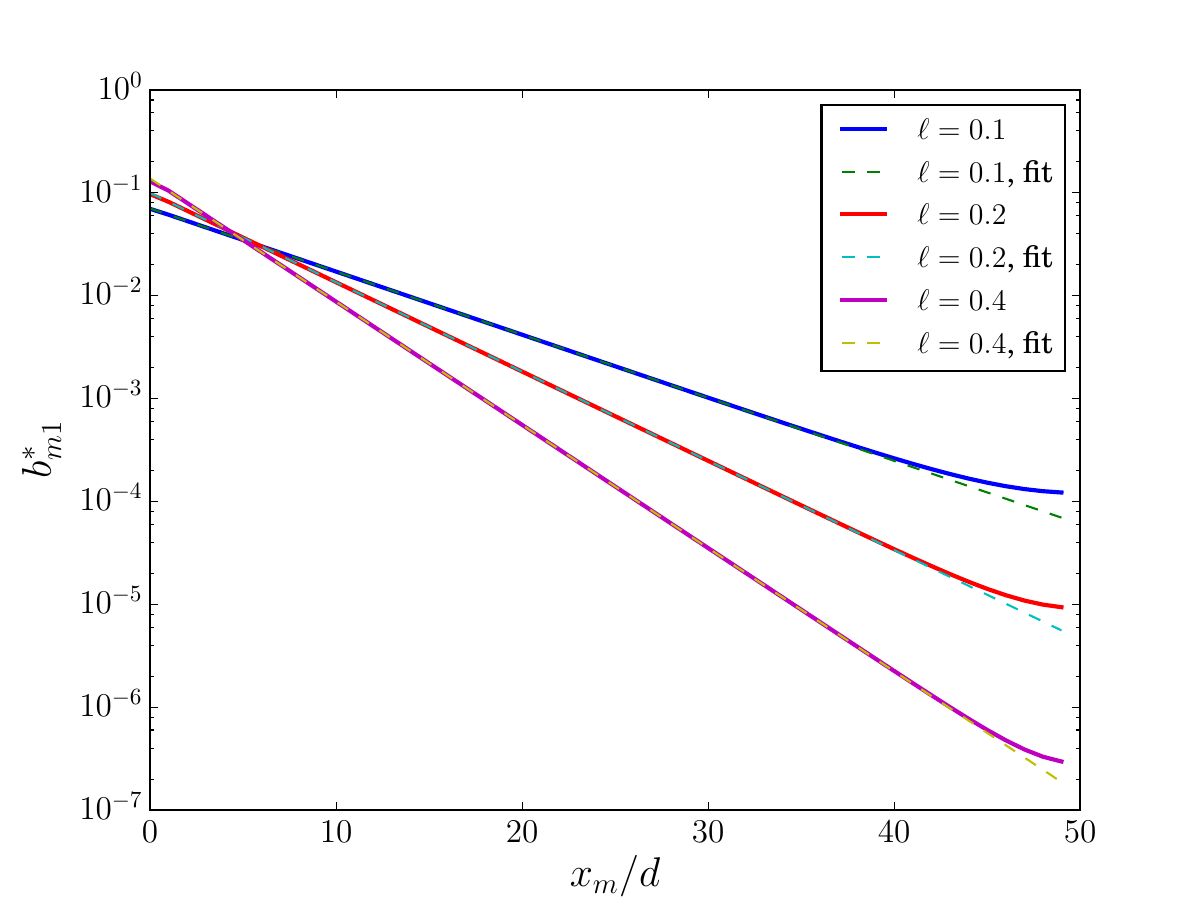}(b)\includegraphics[width=0.35\textwidth]{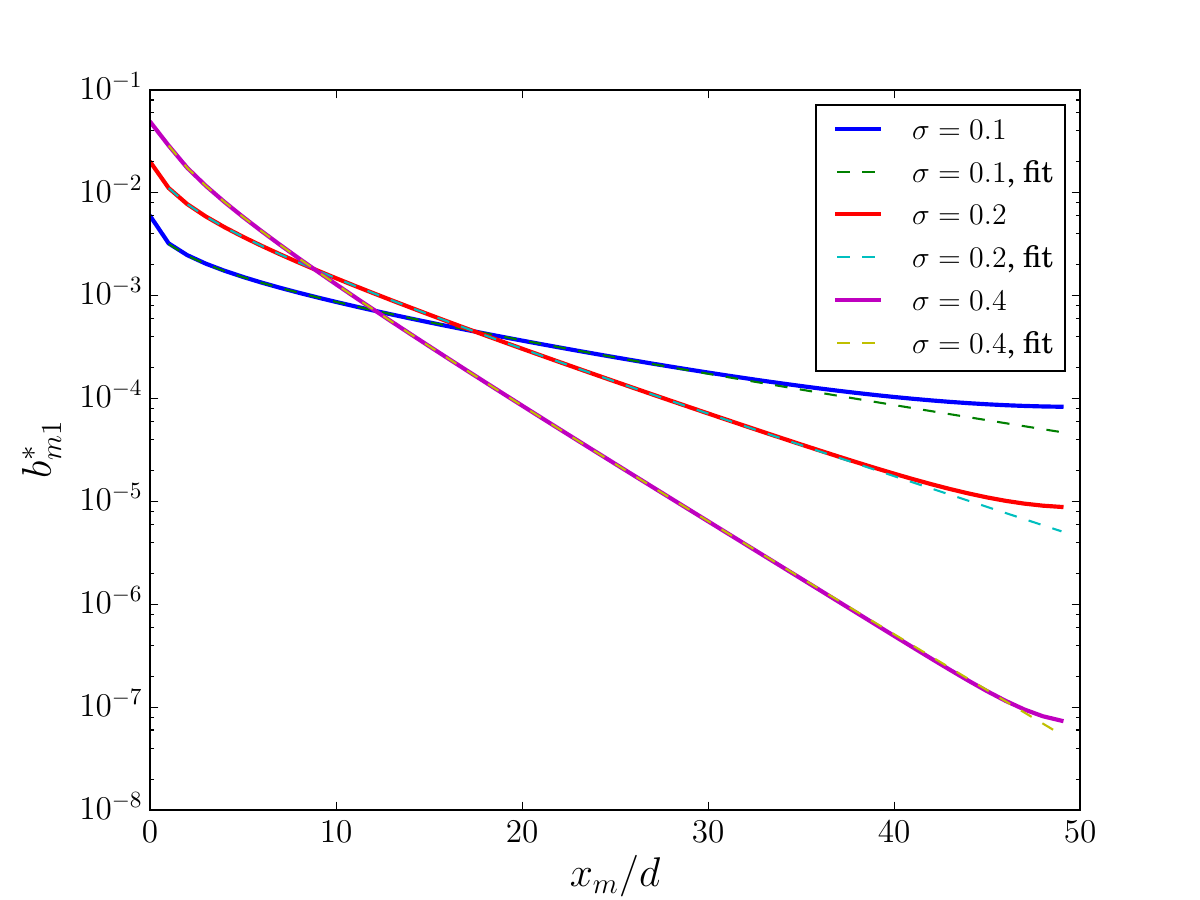}
\par\end{centering}

\begin{centering}
(c)\includegraphics[width=0.35\textwidth]{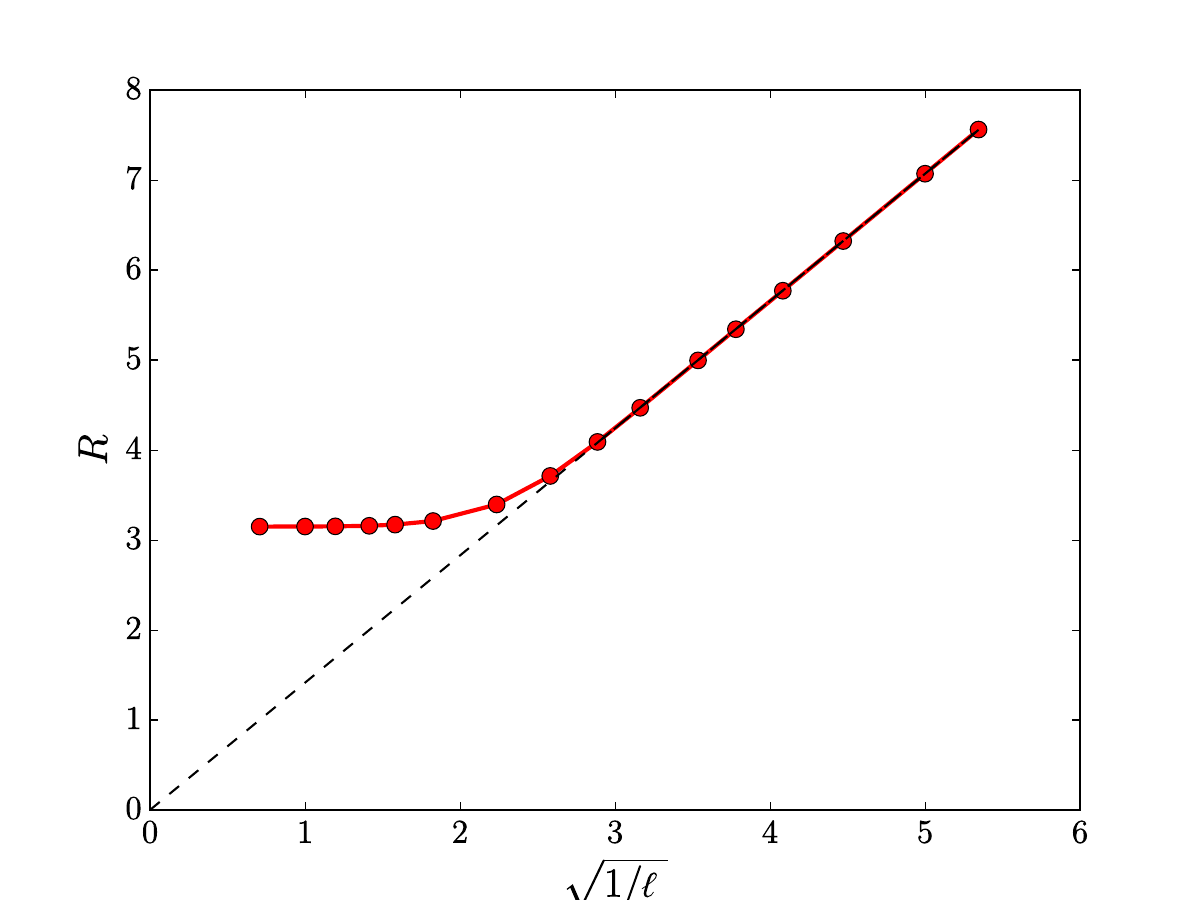}(d)\includegraphics[width=0.35\textwidth]{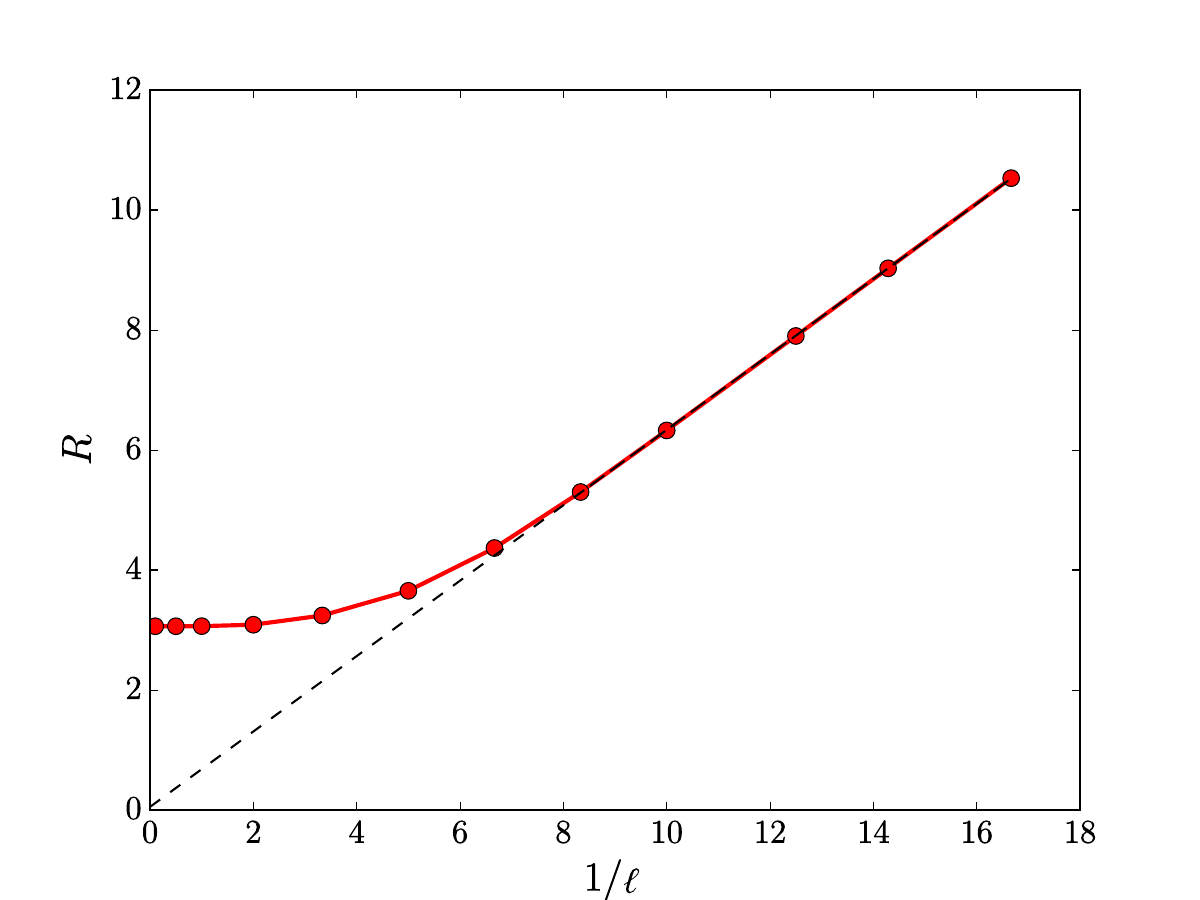}
\par\end{centering}

\centering{}\caption{\label{fig: cmp-kernels} Discrete hopping kernel $G_{ij}$. (a) Comparing
the numerical kernel $b_{m1}^{*}$ with exponential fitting in 1D.
$\kappa=100$, $\Omega=10$, and $\Delta=10$. (b) Comparing $b_{m1}^{*}$
with $K_{0}$ fitting in 2D. (c) The scaling of the hopping radius
$R\sim\ell^{-1/2}$ in 1D versus a length scale $\ell$. (d) The scaling
of the hopping radius $R\sim\ell^{-1}$ in 2D versus $\ell$. }
\end{figure*}

The numerical results fit perfectly for the kernel $G_{D}(r)$ in
both 1D and 2D as shown in Fig. \ref{fig: cmp-kernels}. Moreover,
the predicted hopping radius $R$ fit perfectly with Eq. (\ref{eq: R_eff_hopping_radius})
when the Gaussian $\ell\ll d$ is sufficiently narrow such that the
approximation is good. Both of these fitting give the constant $C_{D}\approx1$.

\section{Chimera patterns in the NLHM}

\subsection{Other initial conditions in 1D and the effect of hopping strength}

\begin{figure}
\begin{centering}

\includegraphics[width=0.9\columnwidth]{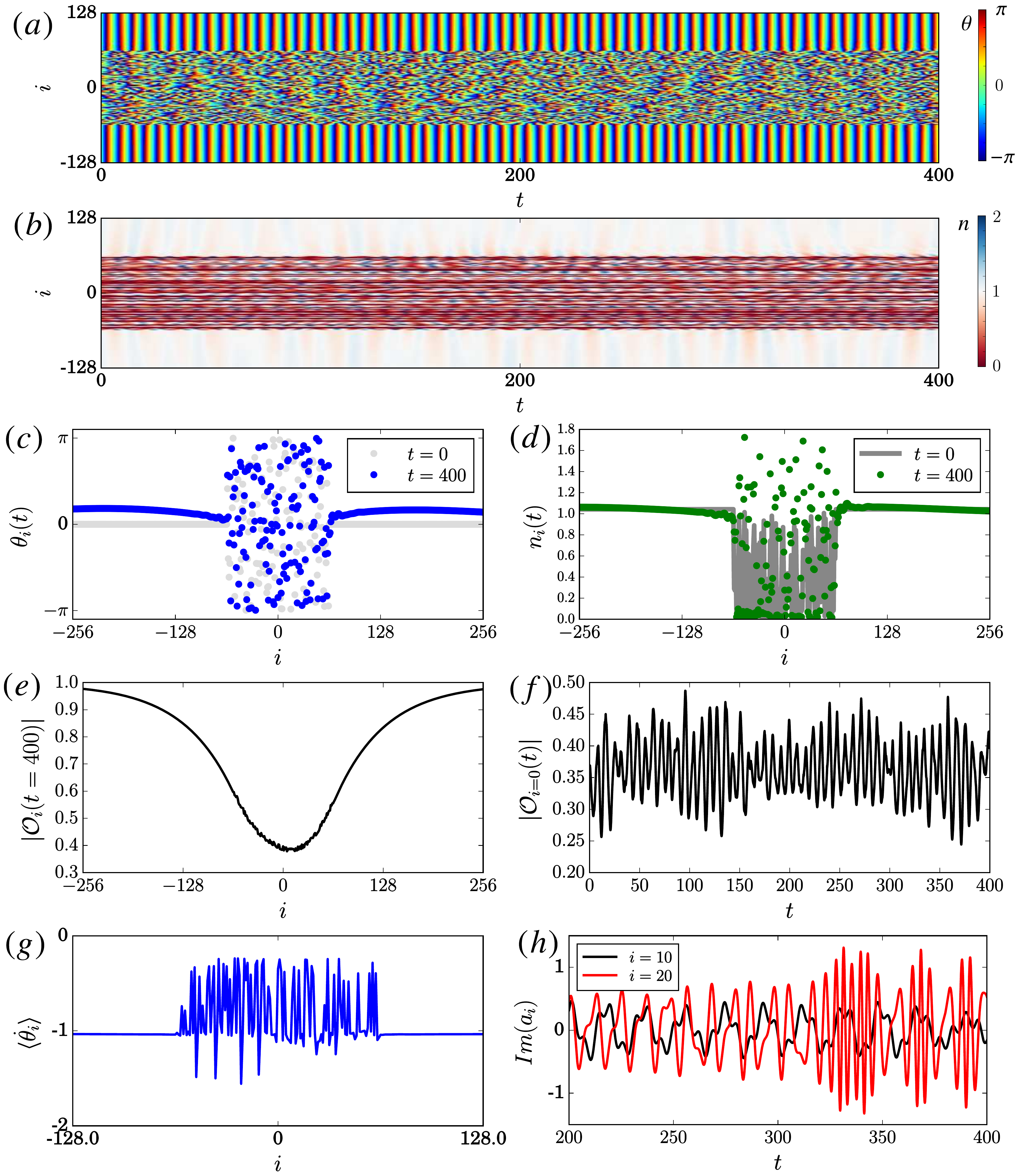}

\par\end{centering}

\centering{}\caption{\label{fig: nlhm1d-randomfull}
Similar to Fig. 2 in the main text, but with initial random phase and random amplitude as given in subfigure (c) and (d).
Note that $|\mathcal{O}_{i=0}|\approx e^{-1}$ for a region to be fully incoherent as calculated in text.
Same parameter as in Fig. 2, but with  $N = \sum_i |a_i|^2 = L$.
}
\end{figure}

As shown in Fig. 2 in the main text, the random phase initial condition with uniform amplitude gives a chimera pattern that has a low local order parameter $|\mathcal{O}|$ at the center.
A lower $|\mathcal{O}|$ can be obtained by using the initial condition that has both random phase and random amplitude as given by Fig. \ref{fig: nlhm1d-randomfull}(c) and \ref{fig: nlhm1d-randomfull}(d).
The magnitude of $|\mathcal{O}|$ is even lower than Fig. 2 in the main text, implying that the state is even less coherent.
Suppose the phase of the region $-R<x<R$ is completely random while constant outside, then the order parameter gives the $|\mathcal{O}_{i=0}| = \sum_{j}G_{ij}e^{i\theta_j} \approx (1/R)\int_R^{\infty} e^{-z/R}dz = e^{-1}$, which is roughly the value for the random phase initial condition at $t=0$ at the center. As shown in Fig. \ref{fig: nlhm1d-randomfull}(e), $|\mathcal{O}_{i=0}|\approx e^{-1}$ at $t=400$, that means the phase are still completely random near the center after some times.

\begin{figure}
\begin{centering}
(a)
\includegraphics[width=0.45\columnwidth]{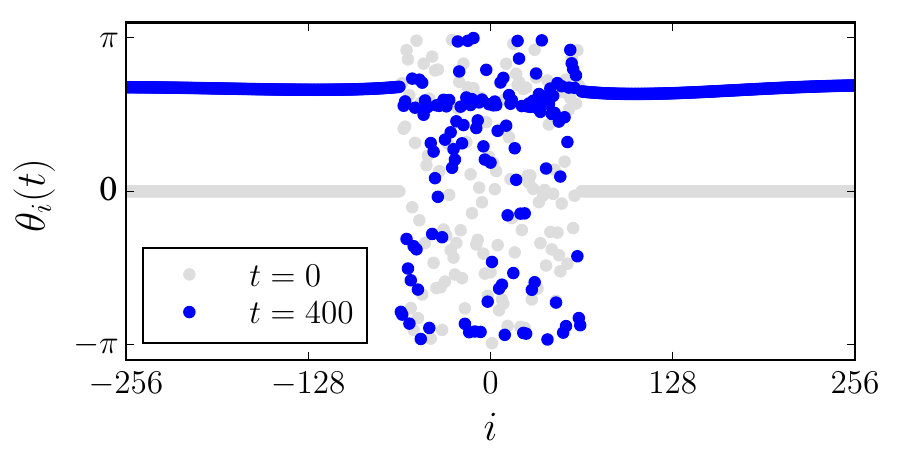}
(b)
\includegraphics[width=0.45\columnwidth]{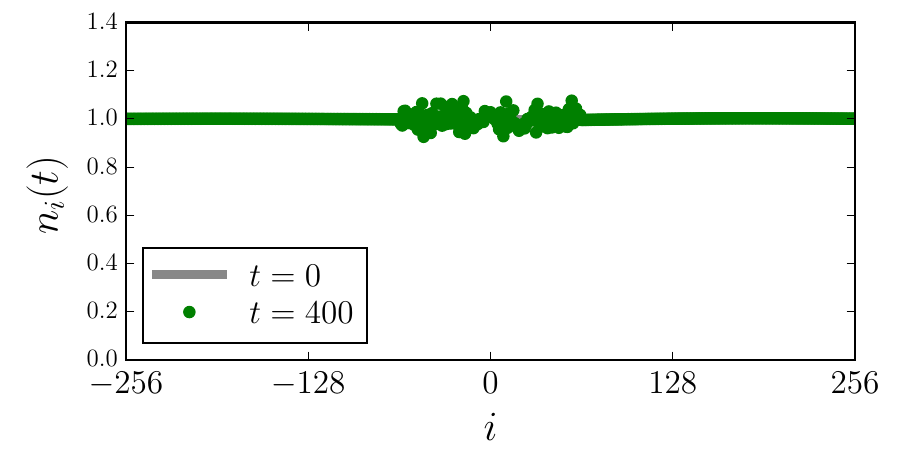}

(c)
\includegraphics[width=0.45\columnwidth]{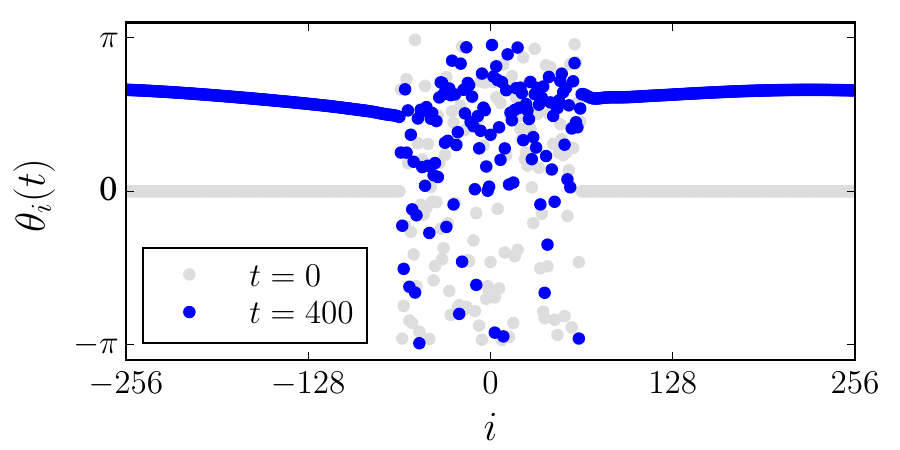}
(d)
\includegraphics[width=0.45\columnwidth]{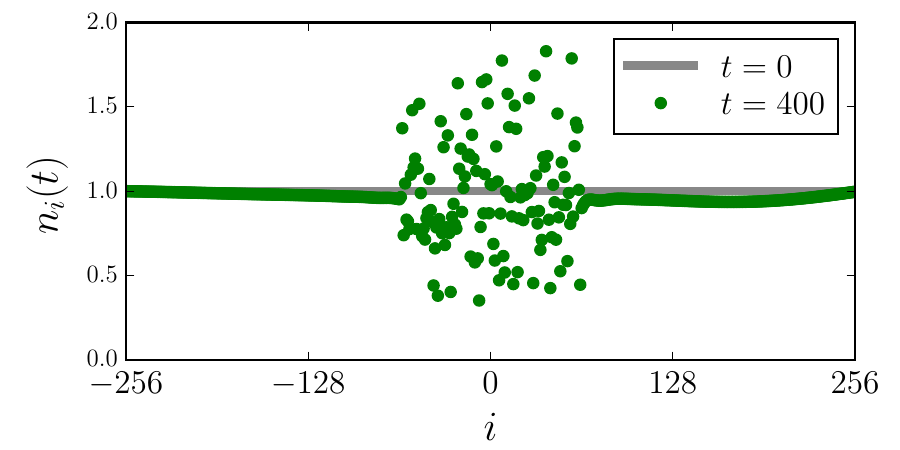}

(e)
\includegraphics[width=0.45\columnwidth]{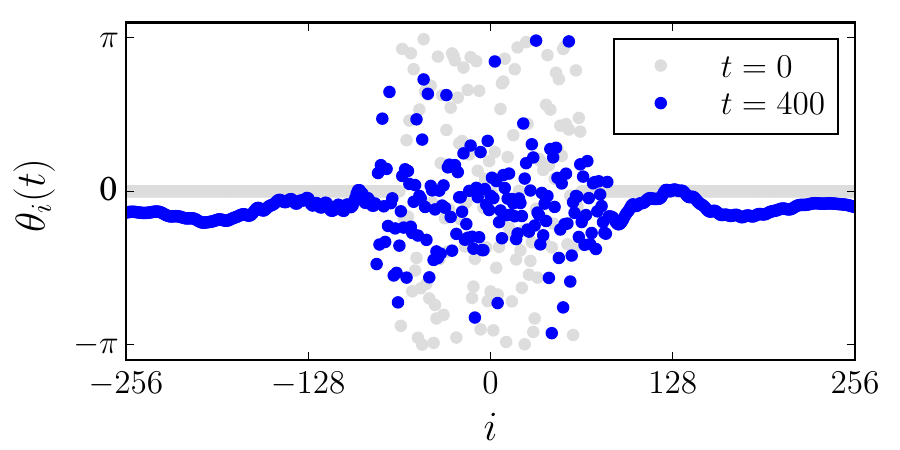}
(d)
\includegraphics[width=0.45\columnwidth]{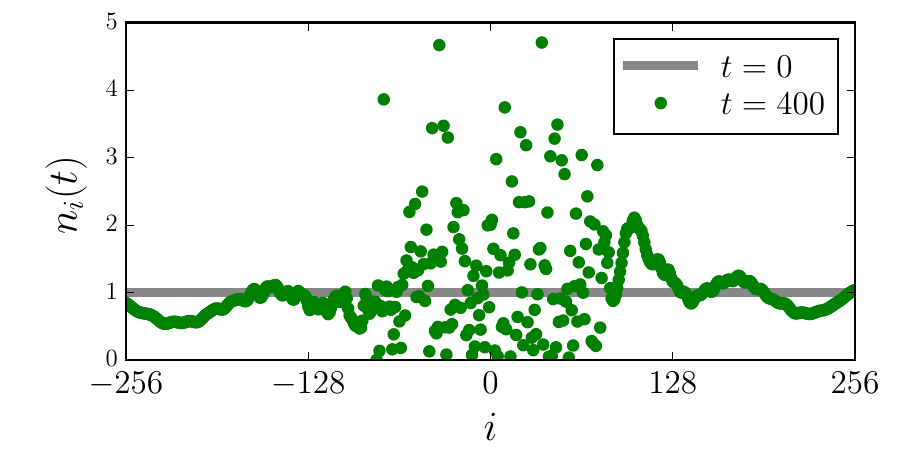}
\par\end{centering}

\centering{}\caption{\label{fig: nlhm1d-hopping-strength}
Effects of different hopping strenght $P$.
(a,b) $P=10^{-4}$,
(c,d) $P=0.1$,
(e,f) $P=5$.
}
\end{figure}

For a sufficiently small hopping strenght $P$, the oscillators are spatially decoupled while the local phase and amplitude are still strongly coupled as shown in Eq. (3) in the main text.
Simulations suggest that stable chimera patterns can still exist even for very small $P$ and in the strong hopping regime as shown in Fig. \ref{fig: nlhm1d-hopping-strength}.
In the regime with very small $P$, the amplitude fluctuates around the initial amplitude by a small amount.
On the other hand, in the strong hopping regime, the chimera pattern with the random phase initial condition still exist, but the amplitude fluctuates a lot larger.
In general, a chimera pattern can be more difficult to emerge from a regular pattern with strong $P$.
For example, with the same vortex initial condition in Fig. 3 in the main text, no incoherence region has formed in the strong hopping regime.
Nevertheless, with the 2D vortex initial condition, the incoherence core can still emerge and it is the one of the main difference observed between 1D and 2D.

\subsection{Spiral wave and vortex-like initial condition}

The snaphots at different time with the spiral initial condition used in
the main text is given in Fig. \ref{fig: nlhm-ic-sipral} (and an
animation). As shown in the figure, the incoherent core is spontaneously
formed near the spatial phase singularity. This dynamic pattern is
essentially invariant under the scaling with fixed $R/L$
as shown in Fig. \ref{fig: nlhm-sipral-R64-L1024}, in which both
system size $L$ and hopping radius $R$ are four times larger. The core
has locally incoherent phase, while the dynamics outside the core
are locally coherent.
This pattern can persist over a long time once formed. So if we start with the random phase core, the chimera pattern can also persist over a long time, as shown in Fig. \ref{fig: nlhm-ic-rvortex}.

Similar to the spiral wave initial condition, the chimera patterns can form with a vortex-like initial condition, as shown in Figs. \ref{fig: nlhm-ic-rvortex} and Figs. \ref{fig: nlhm-ic-vortex}.
The snaphots of the dynamic of a vortex dip at the center used in the main text is given in Fig. \ref{fig: nlhm-ic-vortex}, which can be compared with Fig. \ref{fig: nlhm-ic-sipral}.

The dynamics are very different when the hopping range becomes small
$R\sim d$. With only nearest-neighbors hopping, the system becomes
the discrete Gross-Pitaevskii equation. The difference is very clear
when the system is started from the same random core with nearest-neighbor
hopping as shown in Fig. \ref{fig: nlhm-nn}. The random phase near
the core is a localized perturbation that propagates outward like
a wave and interferes with itself. In this case, no localized chimera
core can be observed.

\subsection{Noise}

\begin{figure}
\begin{centering}
\includegraphics[width=0.5\columnwidth]{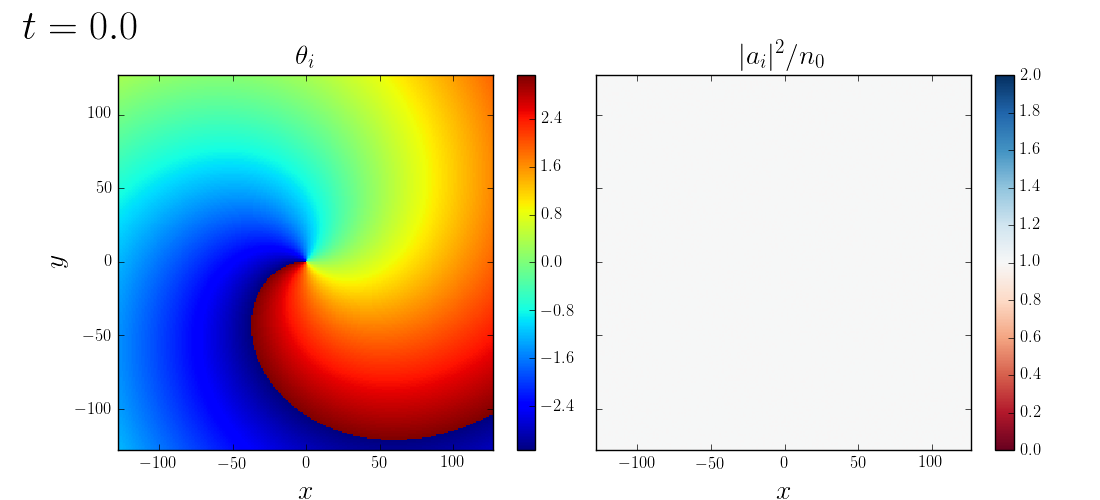}\includegraphics[width=0.5\columnwidth]{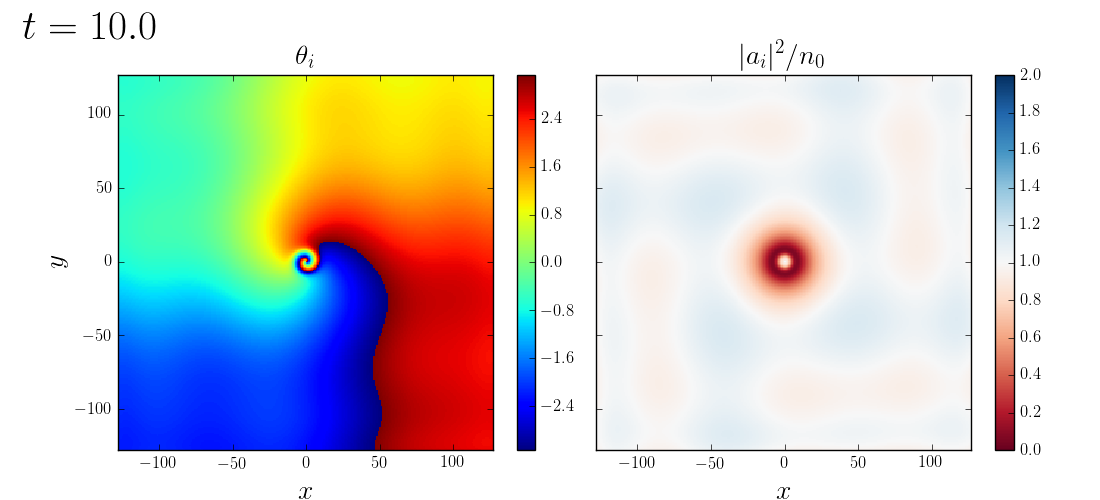}
\par\end{centering}

\begin{centering}
\includegraphics[width=0.5\columnwidth]{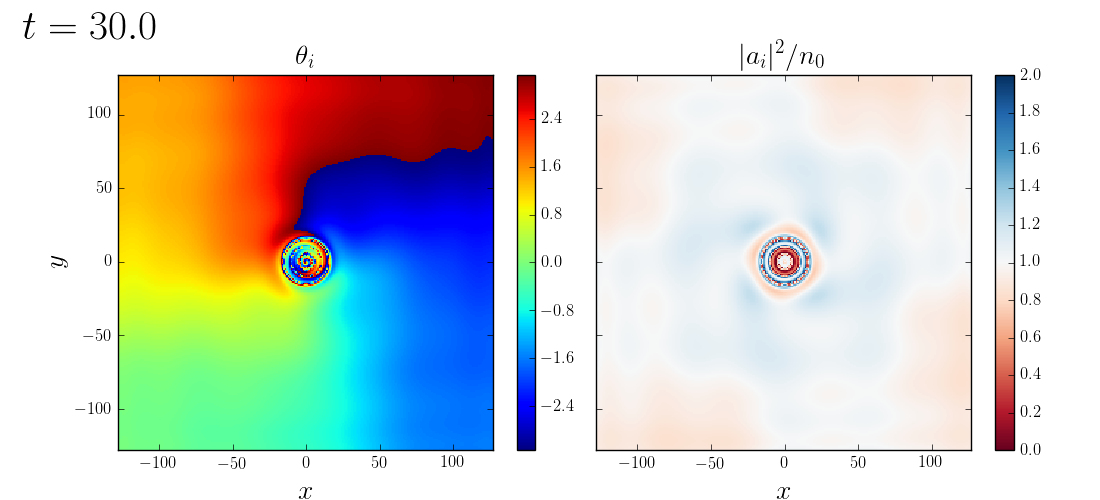}\includegraphics[width=0.5\columnwidth]{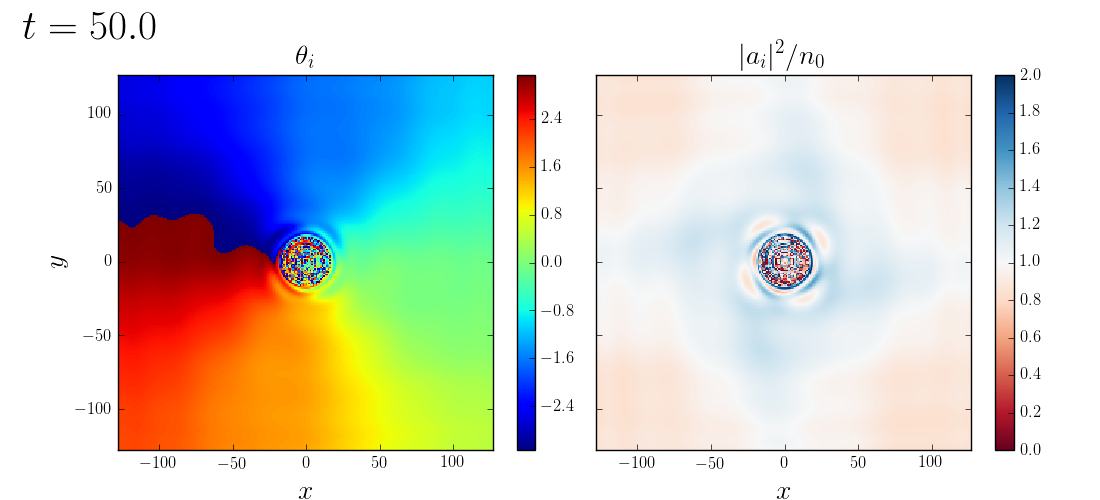}
\par\end{centering}

\begin{centering}
\includegraphics[width=0.5\columnwidth]{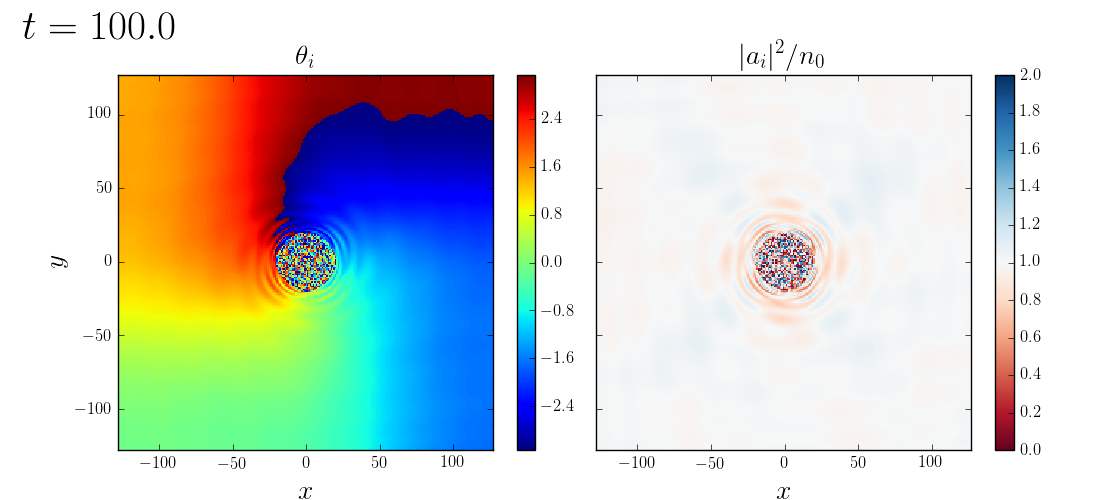}\includegraphics[width=0.5\columnwidth]{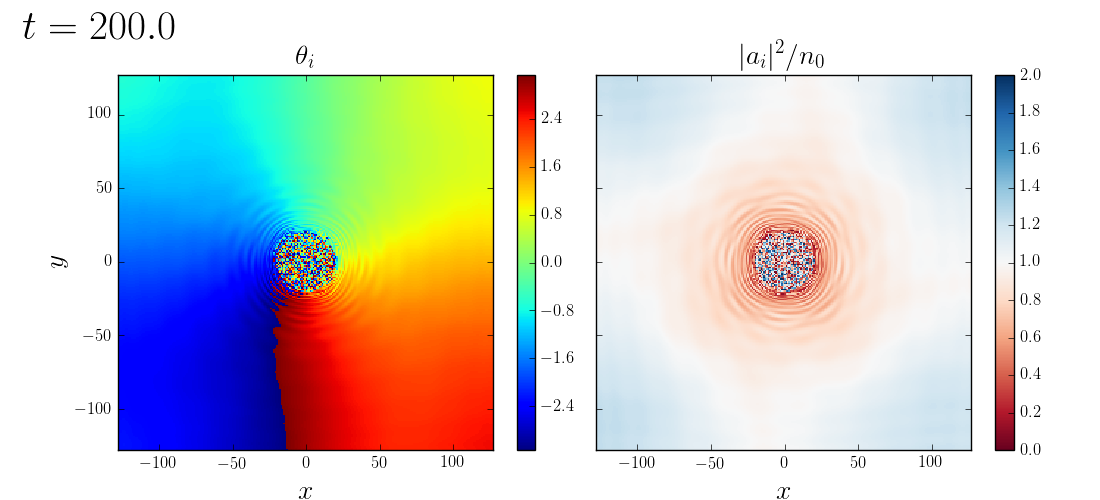}
\par\end{centering}

\caption{\label{fig: nlhm-ic-sipral} Time evolution of the initial spiral
with $k_{s}=0.01$. We used nonlocal hopping $P/(Un_{0})=0.5$ and
$R=16$ in a system with a lattice size of $L=256$.}
\end{figure}

\begin{figure}
\begin{centering}
\includegraphics[width=1\columnwidth]{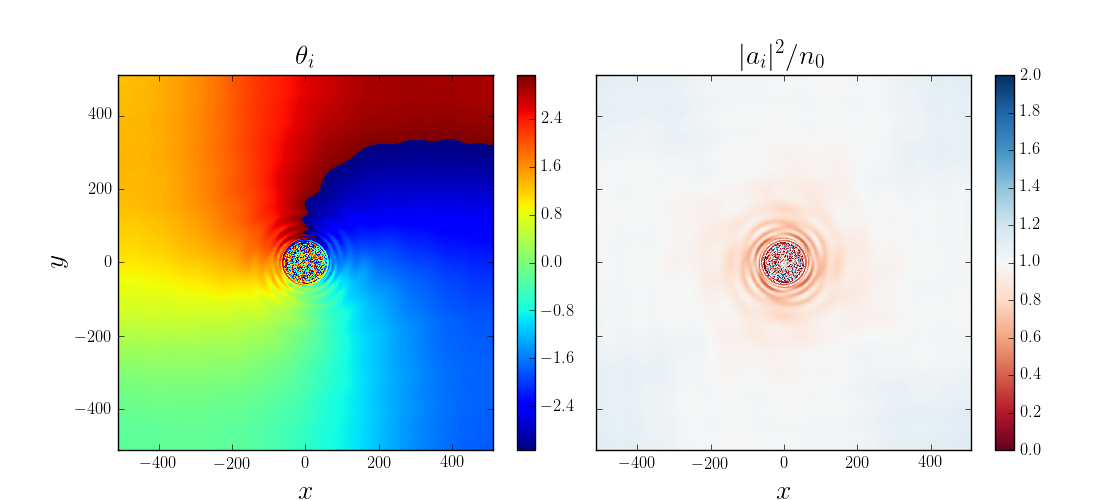}
\par\end{centering}

\caption{\label{fig: nlhm-sipral-R64-L1024} Similar to $t=100$ Fig. \ref{fig: nlhm-ic-sipral}
but in a larger system with initial spiral $k_{s}=0.0025$. We use
nonlocal hopping $P/(Un_{0})=0.5$ and $R=64$ in a system with a
lattice size of $L=1024$. This can be compared with Fig. \ref{fig: nlhm-ic-sipral}e.}
\end{figure}

\begin{figure}
\begin{centering}
\includegraphics[width=0.5\columnwidth]{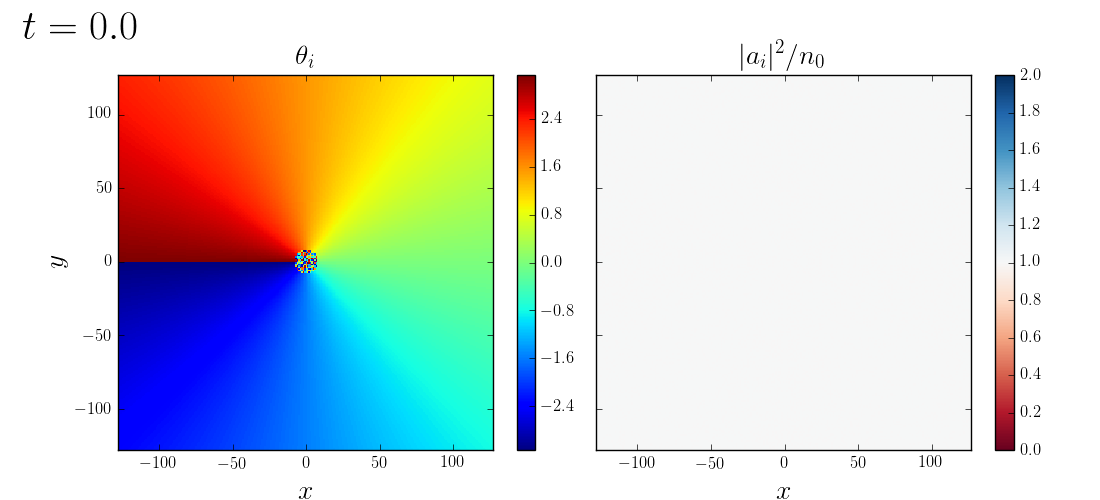}\includegraphics[width=0.5\columnwidth]{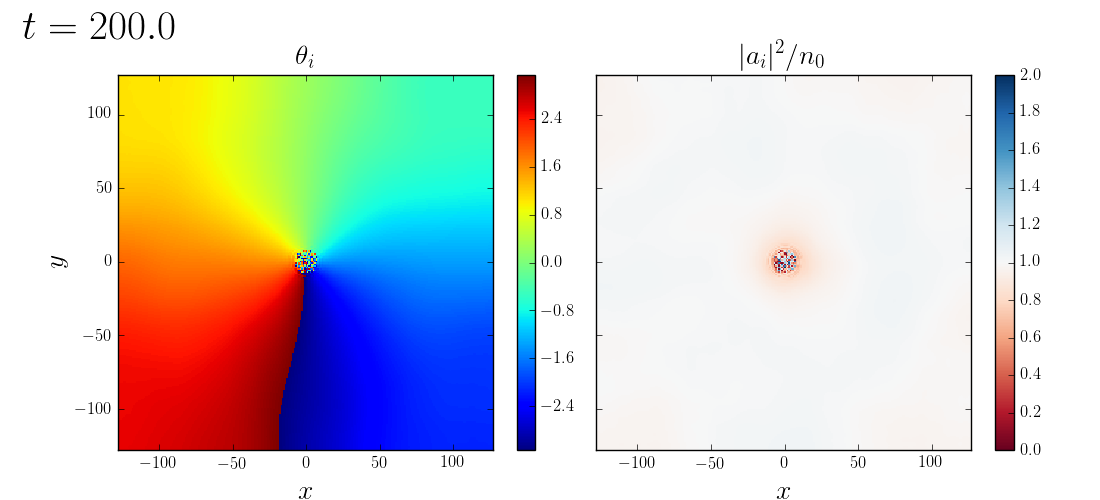}
\par\end{centering}

\caption{\label{fig: nlhm-ic-rvortex} Time evolution of an initial random phase core with radius $R_{core}=8$ and uniform amplitude. $P/(Un_{0})=0.5$, $R=8$, and $L=256$.}
\end{figure}

\begin{figure}
\begin{centering}
\includegraphics[width=0.5\columnwidth]{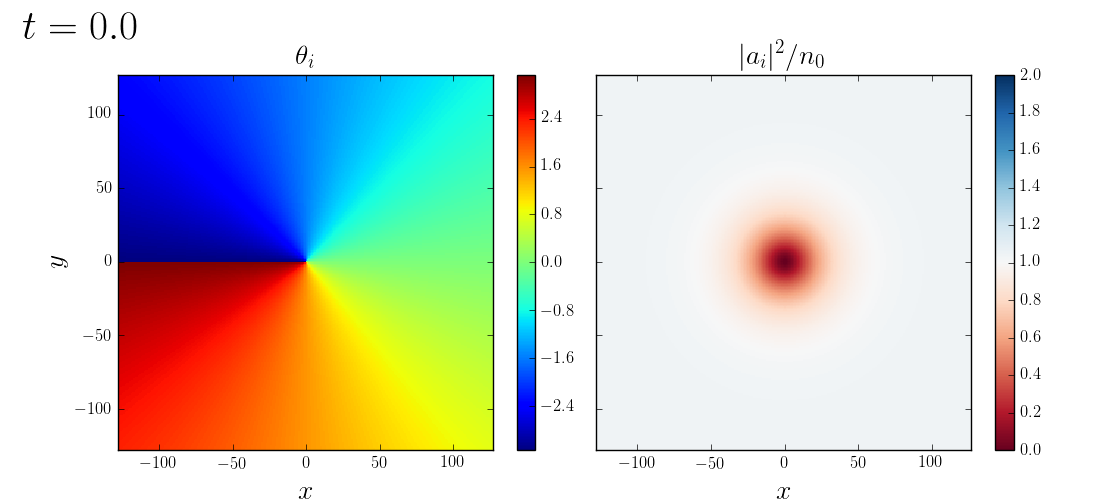}\includegraphics[width=0.5\columnwidth]{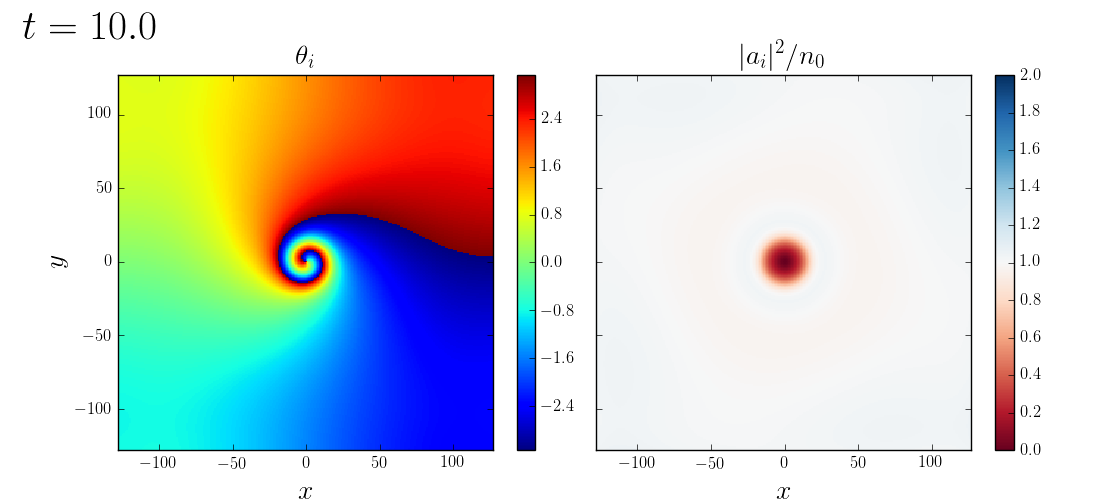}
\par\end{centering}

\begin{centering}
\includegraphics[width=0.5\columnwidth]{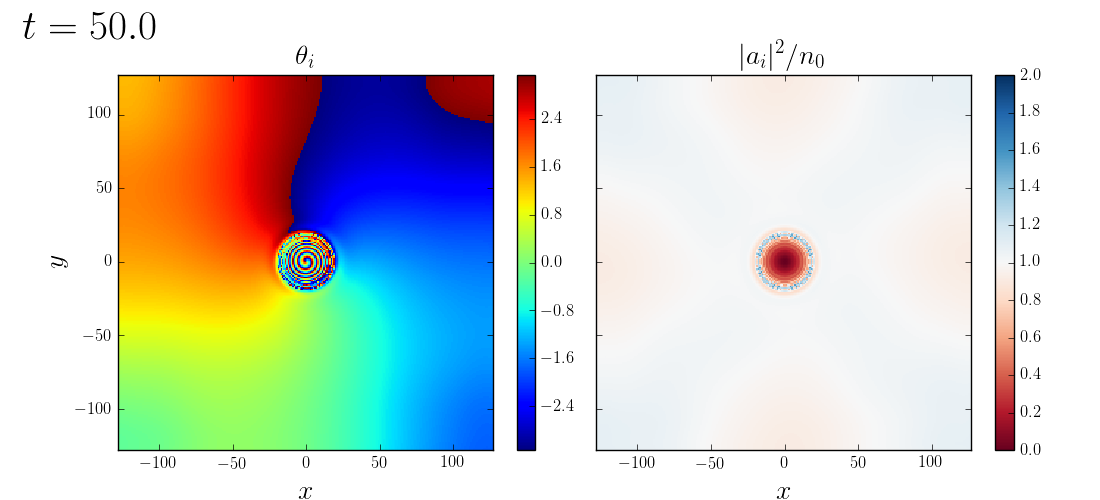}\includegraphics[width=0.5\columnwidth]{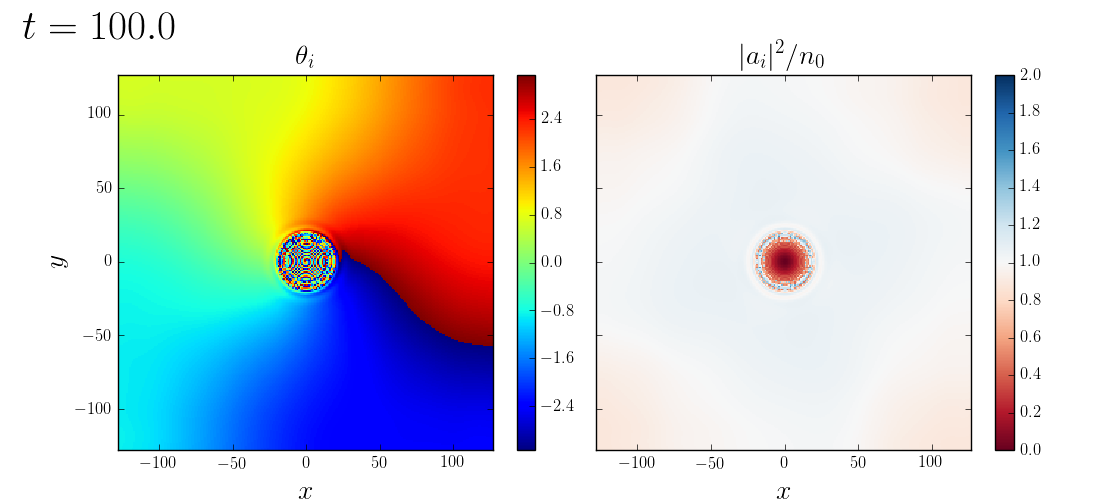}
\par\end{centering}

\begin{centering}
\includegraphics[width=0.5\columnwidth]{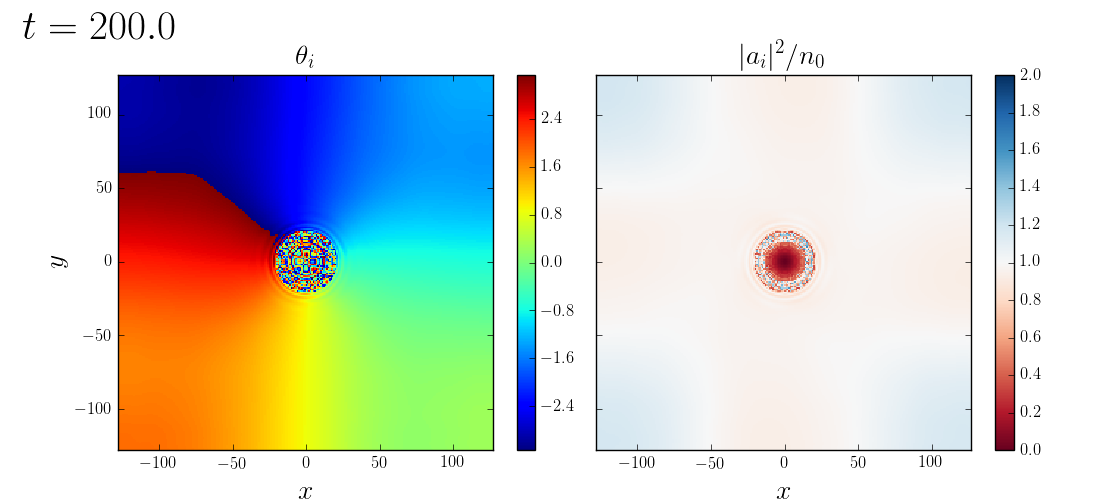}\includegraphics[width=0.5\columnwidth]{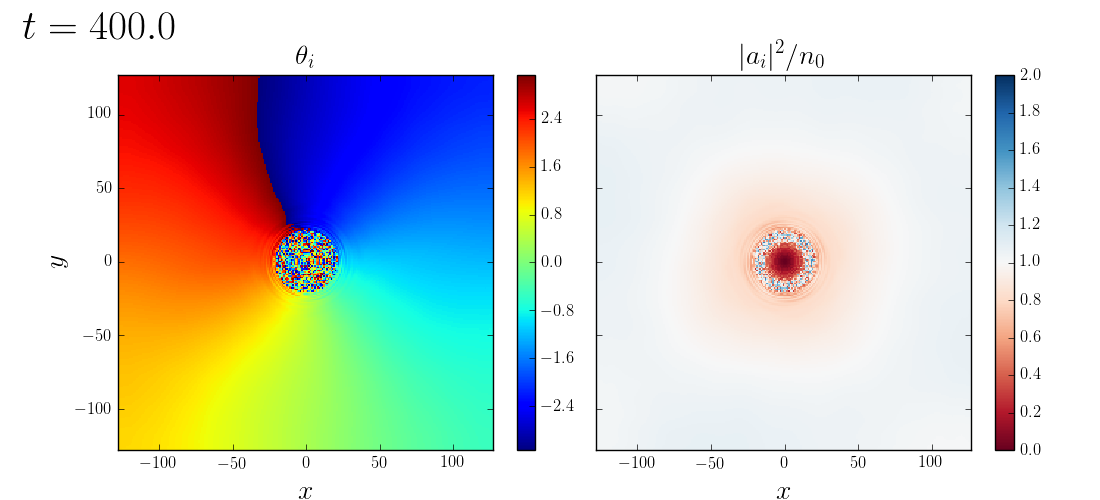}
\par\end{centering}

\caption{\label{fig: nlhm-ic-vortex} Time evolution of an initial vortex
with $R_{vortex}=16$. We used nonlocal hopping $P/(Un_{0})=0.1$ and
$R=16$ in a system with a lattice size of $L=256$.}
\end{figure}

\begin{figure}
\begin{centering}
\includegraphics[width=0.5\columnwidth]{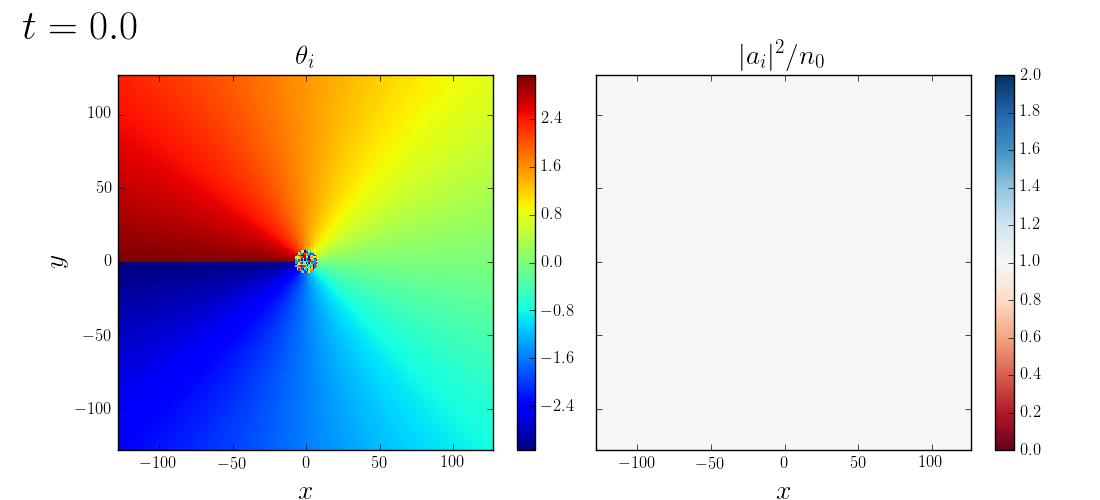}\includegraphics[width=0.5\columnwidth]{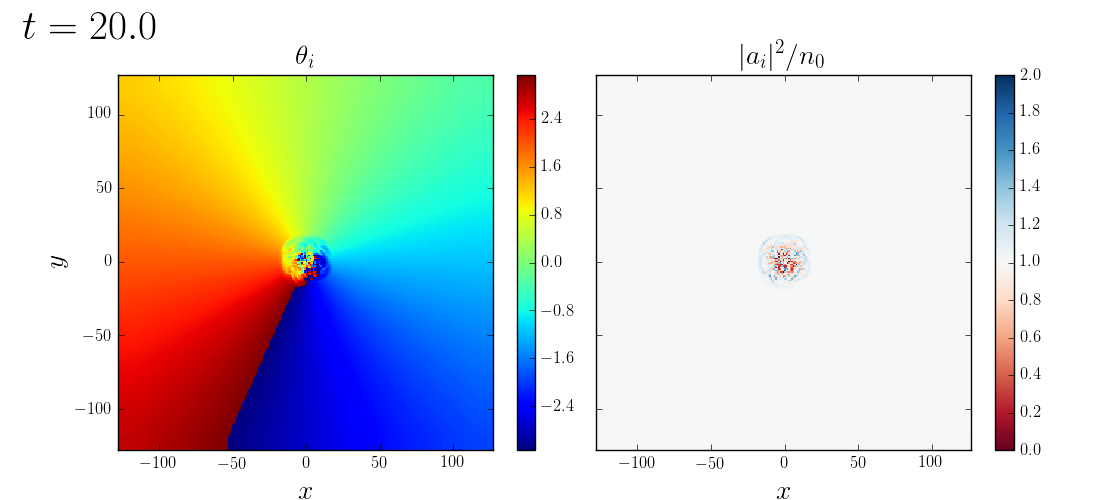}
\par\end{centering}

\begin{centering}
\includegraphics[width=0.5\columnwidth]{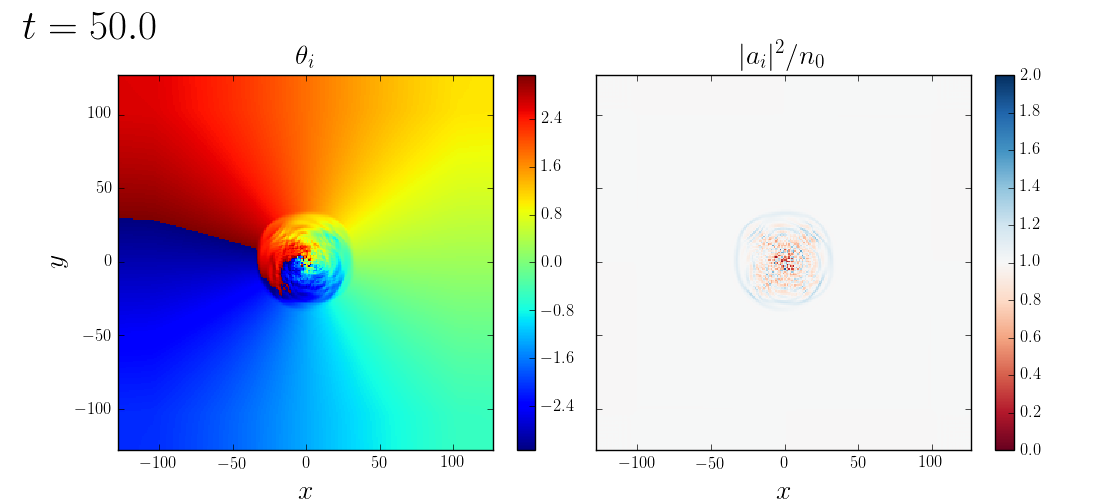}\includegraphics[width=0.5\columnwidth]{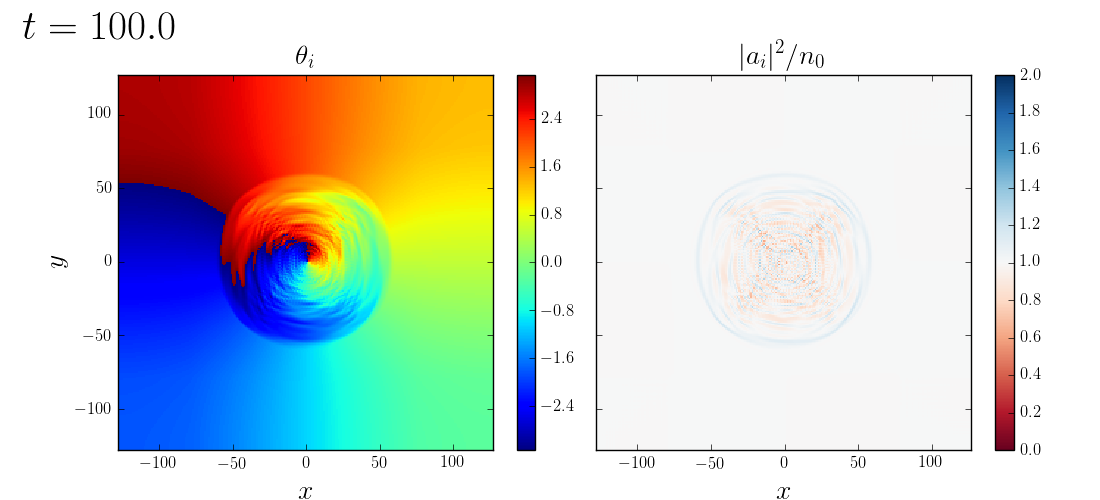}
\par\end{centering}

\begin{centering}
\includegraphics[width=0.5\columnwidth]{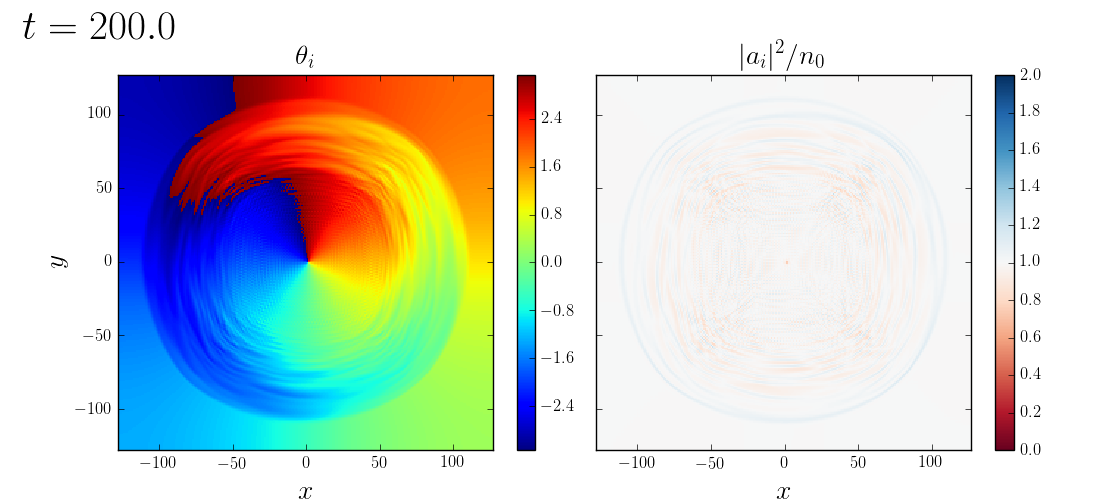}\includegraphics[width=0.5\columnwidth]{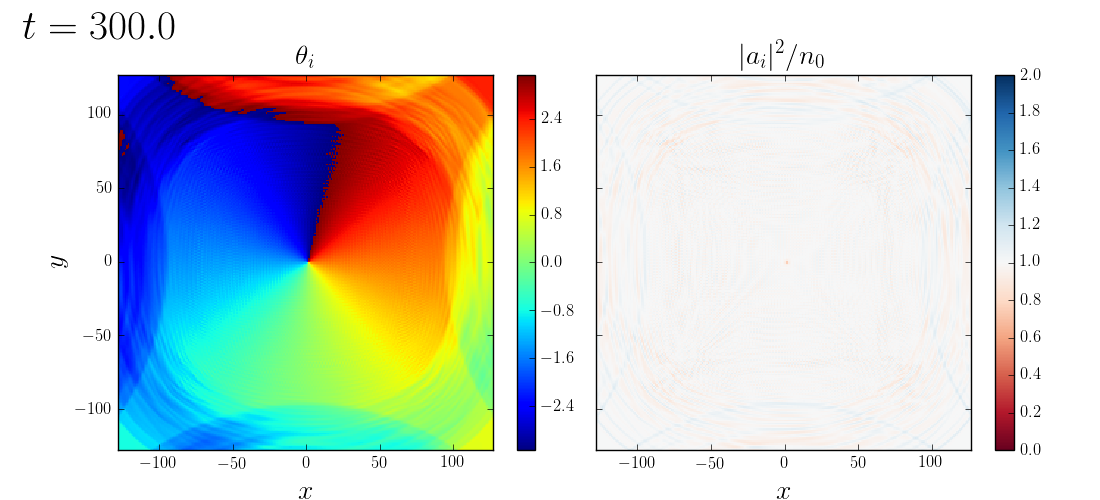}
\par\end{centering}

\caption{\label{fig: nlhm-nn} Time evolution of an initial random core with radius $R_{core}=8$ with
only nearest-neighbor hopping. $P/(Un_{0})=0.5$ and $L=256$.}
\end{figure}

\begin{figure}
\begin{centering}
(a)
\includegraphics[width=0.5\columnwidth]{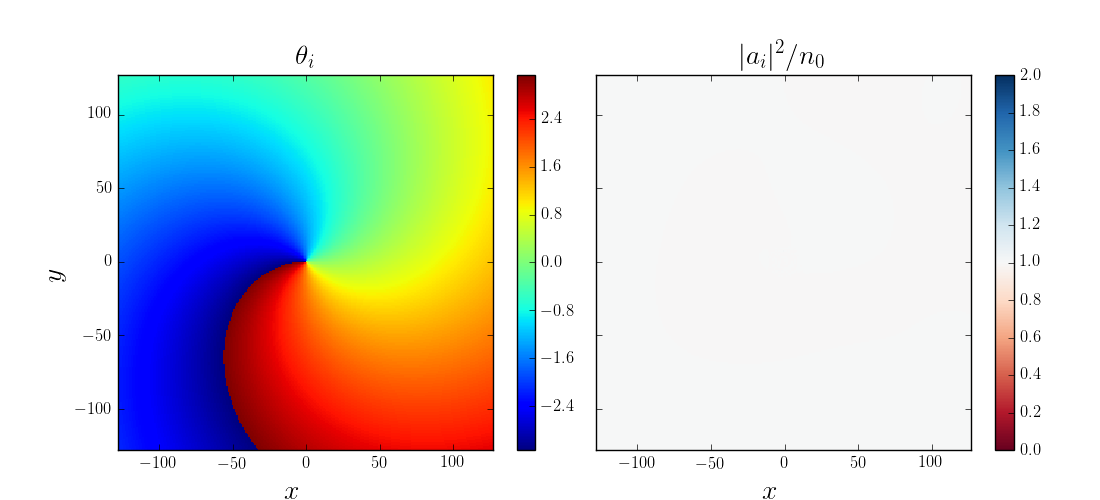}

(b)
\includegraphics[width=0.5\columnwidth]{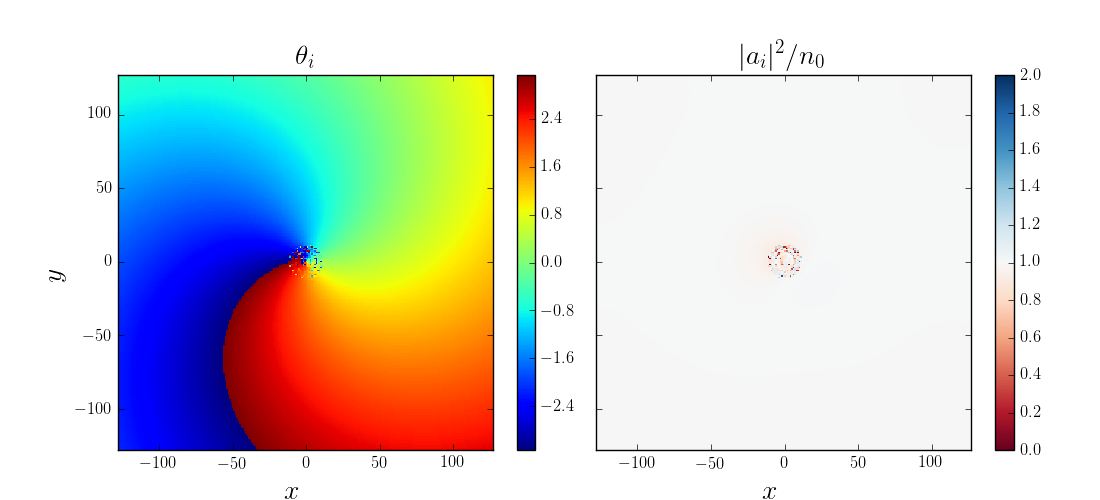}

(c)
\includegraphics[width=0.5\columnwidth]{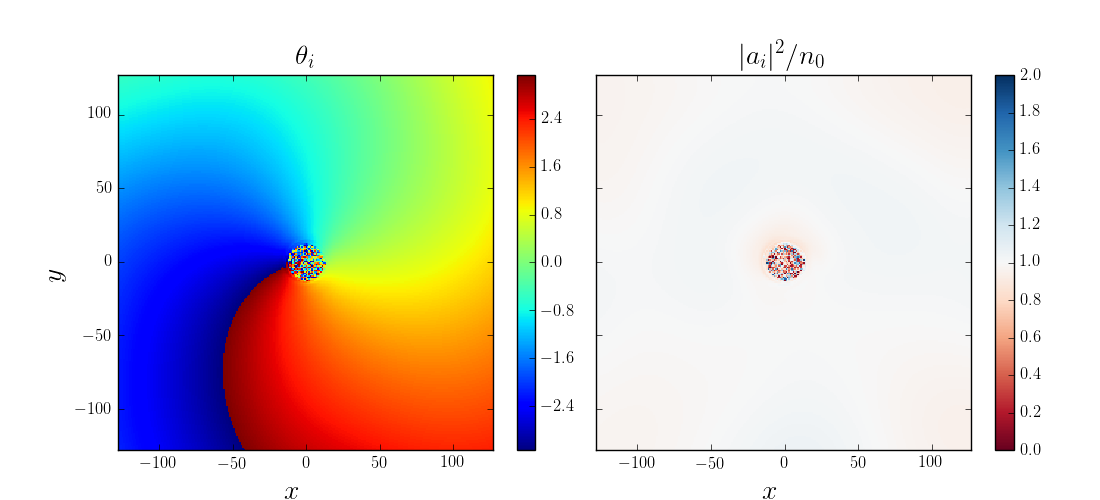}
\par\end{centering}

\caption{\label{fig: nlhm-time-reversal} Backward time propagation for the duration $t=200$ using the
state $t=200$ in Fig. \ref{fig: nlhm-ic-sipral} as the initial condition.
(a) No noise,
(b) $\chi_{noise}=10^{-11}$,
(c) $\chi_{noise}=10^{-10}$, where the single-shot noises are add
before the backward propagation. The noise added is very tiny $\chi_{noise}/|a_{i}|\sim\chi_{noise}$
since $|a_{i}|\sim1$ is used.}
\end{figure}

The behaviour of the core is very different from the background. In
particular, the dynamics of the core are very sensitive to small fluctuations
or noises. For example, we can evolve the system backward in time
and expect it to go back to the initial spiral, as shown in \ref{fig: nlhm-time-reversal}a,
if the system starts from the state in Fig. \ref{fig: nlhm-ic-sipral}f.
The sensitivity of the core region can be tested by adding a small
single-shot noise
\begin{equation}
a_{i}\to a_{i}+\chi_{\text{noise}}\xi_{i},
\end{equation}
where the noise is Gaussian with $\langle\xi_{i}\rangle=0$, $\langle\text{Re}(\xi_{i})\text{Re}(\xi_{i'})\rangle=\delta_{i,i'}$,
and $\langle\text{Im}(\xi_{i})\text{Im}(\xi_{i'})\rangle=\delta_{i,i'}$,
with amplitude $\chi_{\text{noise}}$. This noise can be added to
the state in Fig. \ref{fig: nlhm-ic-sipral}f as a perturbation before
the backward propagation. As shown in Fig. \ref{fig: nlhm-time-reversal}b
and \ref{fig: nlhm-time-reversal}c, the system cannot go back to
the spiral even with a noise as low as $\chi_{\text{noise}}=10^{-11}$,
as compared to the order $\mathcal{O}(1)$ of the amplitude and phase.
This suggests that the core region is very sensitive to the initial
condition. This is in stark contrast to with the behaviour of the
coherent background, which can go back to the same local states as
in the noiseless case.

\subsection{Loss}

\begin{figure}
\begin{centering}
\includegraphics[width=0.49\columnwidth]{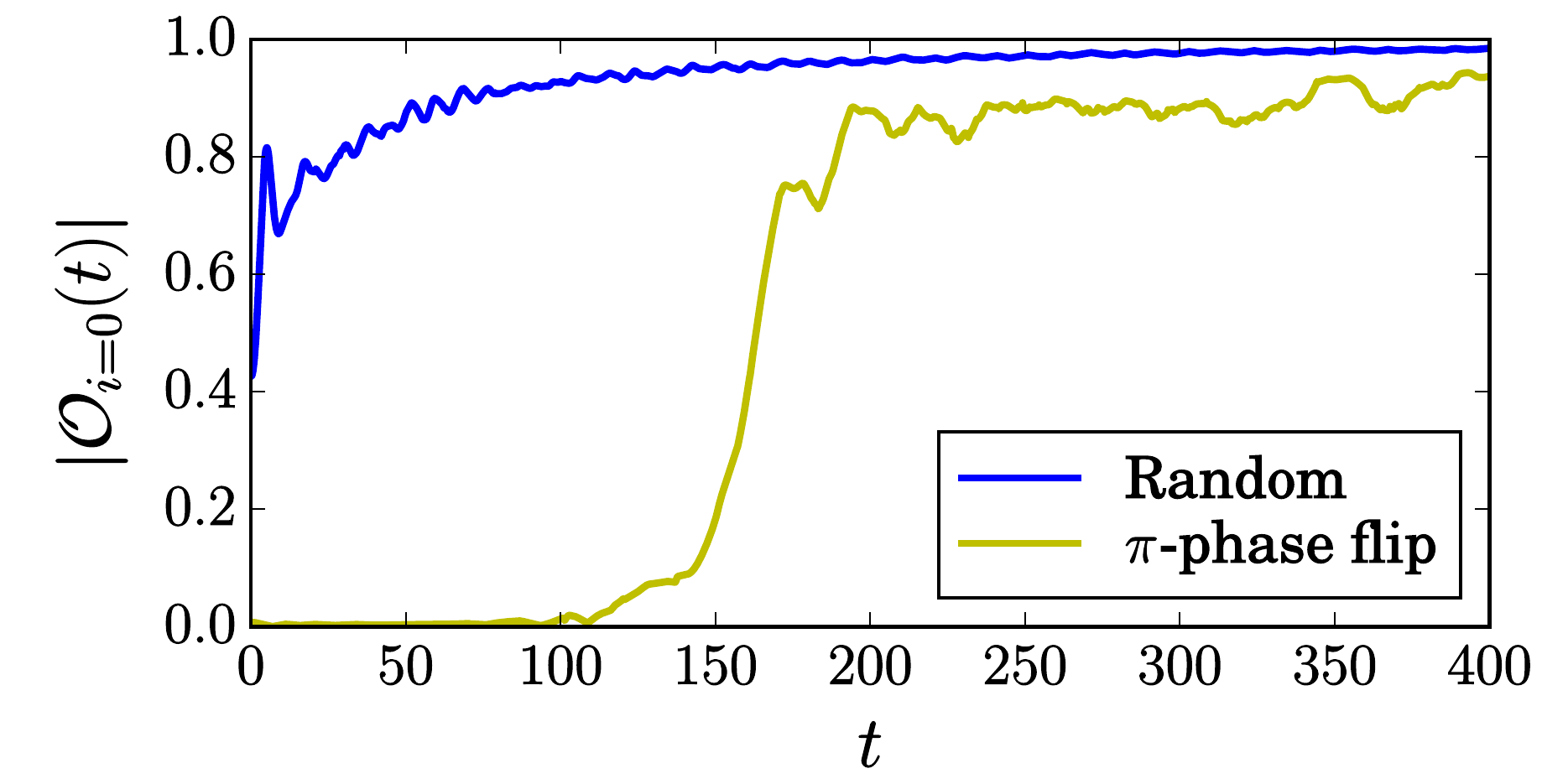}
\par\end{centering}

\centering{}\caption{\label{fig: nlhm1d-loss}
NLHM in 1D under nonlinear loss for the two initial conditions in Fig. 2 and Fig. 3 in the main text. The loss used is $U \to U-iU_{loss}$ where $U_{loss}=0.01$.
The center becomes synchronized with the system when $\mathcal{O}\sim 1$. 
}
\end{figure}

\begin{figure}
\begin{centering}
\subfloat[$U_{loss}=0.001$, $t=60$]{\begin{centering}
\includegraphics[width=0.27\columnwidth]{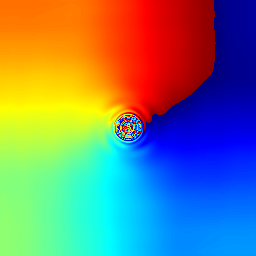}
\par\end{centering}

} \subfloat[$U_{loss}=0.001$, $t=100$]{\begin{centering}
\includegraphics[width=0.27\columnwidth]{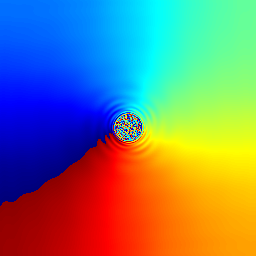}
\par\end{centering}

}
\par\end{centering}

\begin{centering}
\subfloat[$U_{loss}=0.005$, $t=60$]{\begin{centering}
\includegraphics[width=0.27\columnwidth]{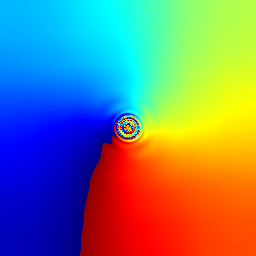}
\par\end{centering}

} \subfloat[$U_{loss}=0.005$, $t=100$]{\begin{centering}
\includegraphics[width=0.27\columnwidth]{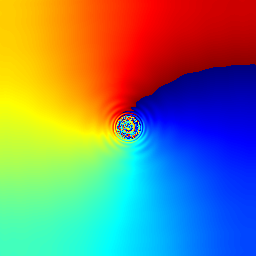}
\par\end{centering}

}
\par\end{centering}

\begin{centering}
\subfloat[$U_{loss}=0.01$, $t=60$]{\begin{centering}
\includegraphics[width=0.27\columnwidth]{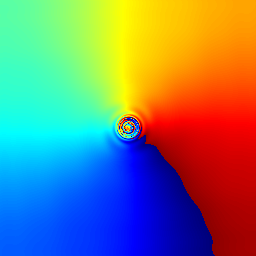}
\par\end{centering}

} \subfloat[$U_{loss}=0.01$, $t=100$]{\begin{centering}
\includegraphics[width=0.27\columnwidth]{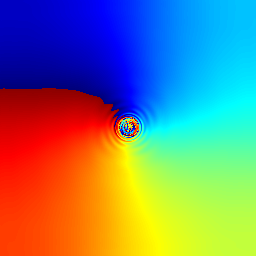}
\par\end{centering}

}
\par\end{centering}

\begin{centering}
\subfloat[$U_{loss}=0.02$, $t=60$]{\begin{centering}
\includegraphics[width=0.27\columnwidth]{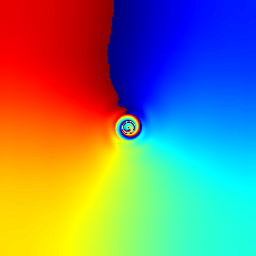}
\par\end{centering}

} \subfloat[$U_{loss}=0.02$, $t=100$]{\begin{centering}
\includegraphics[width=0.27\columnwidth]{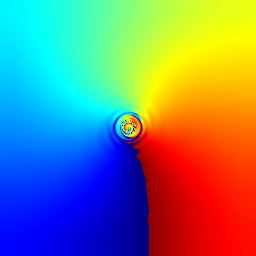}
\par\end{centering}

}
\par\end{centering}

\caption{\label{fig: nlhm-loss} Similar to Fig. \ref{fig: nlhm-ic-sipral}
with loss $U_{loss}=0.001,0.005,0.01,0.02$ from top to bottom. (left) $t=60$, (right)
$t=100$. Parameters: $U=1$, $n_{0}=1$, $P=0.5$, $R=16$, and $L=256$.}
\end{figure}

The nonlinear particle loss can be modelled by the replacing $U\to U-iU_{\text{loss}}$.
Therefore, the dynamic equation with loss is 

\begin{equation}
i\hbar\dot{a}_{i}=(U-iU_{\text{loss}})|a_{i}|^{2}a_{i}-P\sum_{j}G_{ij}a_{j}.\label{eq: NLHM-dynamic-with-loss}
\end{equation}
The particle loss can be calculated by the time-derivative of the
total number of particles:
\begin{equation}
\frac{dN}{dt}=\frac{d}{dt}\sum_{i}|a_{i}|^{2}=\sum_{i}\left(\dot{a}_{i}^{*}a_{i}+a_{i}^{*}\dot{a_{i}}\right)=\sum_{i}-\left(2\frac{U_{\text{loss}}}{\hbar}\right)|a_{i}|^{4},
\end{equation}
where the dynamic equations for $\dot{a}_{i}$ and $\dot{a}_{i}^{*}$
are substituted above. Suppose all lattice sites have the same number
of particles $|a_{i}|^{2}=n$, then the equation becomes
\begin{equation}
\frac{dn}{dt}=-\left(2\frac{U_{\text{loss}}}{\hbar}\right)n^{2}.
\end{equation}
Let $\chi=2U_{loss}/\hbar$ and then by solving the equation, we have
\begin{equation}
n(t)=\frac{1}{n_{0}^{-1}+\chi t},
\end{equation}
where $n_{0}$ is the initial number of particles per site. Suppose
$n_{0}$ is very large, then the time for half of the particle to
become lost is $\chi t=2$ or $U_{loss}t/\hbar=1$.
As shown by the order parameter $|\mathcal{O}_i|$ in Fig. \ref{fig: nlhm1d-loss}, the system becomes more and more coherent over time with loss.
Moreover, as shown in Fig. \ref{fig: nlhm-loss}, the phase of the chimera patterns changes slightly compared to the lossless case on timescales where many more than half of the particles remain in the system.

\end{document}